\documentclass[fleqn,usenatbib]{mnras}

\usepackage{newtxtext,newtxmath}

\usepackage[T1]{fontenc}

\DeclareRobustCommand{\VAN}[3]{#2}
\let\VANthebibliography\thebibliography
\def\thebibliography{\DeclareRobustCommand{\VAN}[3]{##3}\VANthebibliography}

\usepackage{tablefootnote}

\usepackage{graphicx}	
\usepackage{amsmath}	

\title[Magnetized Multiple Formation]{Fragmentation of Dense Rotation-Dominated Structures Fed by Collapsing Gravo-magneto-Sheetlets and Origin of Misaligned 100 au-Scale Binaries and Multiple Systems}

\author[Y. Tu et al.]{
Yisheng Tu,$^{1}$\thanks{E-mail: yt2cr@virginia.edu}
Zhi-Yun Li,$^{1}$
Zhaohuan Zhu,$^{2}$
Chun-Yen Hsu,$^{1}$
\\
$^{1}$Astronomy Department, University of Virginia, Charlottesville, VA 22904, USA\\
$^{2}$Department of Physics and Astronomy, University of Nevada, Las Vegas, NV, 89154-4002, USA}

\date{Accepted XXX. Received YYY; in original form ZZZ}

\pubyear{2015}

\begin{document}
\label{firstpage}
\pagerange{\pageref{firstpage}--\pageref{lastpage}}
\maketitle

\begin{abstract}

The majority of stars are in binary/multiple systems. How such systems form in turbulent, magnetized cores of molecular clouds in the presence of non-ideal MHD effects remains relatively under-explored. Through ATHENA++-based non-ideal MHD AMR simulations with ambipolar diffusion, we show that the collapsing protostellar envelope is dominated by dense gravo-magneto-sheetlets, a turbulence-warped version of the classic pseudodisk produced by anisotropic magnetic resistance to the gravitational collapse, in agreement with previous simulations of turbulent, magnetized single-star formation. The sheetlets feed mass, magnetic fields, and angular momentum to a Dense ROtation-Dominated (DROD) structure, which fragments into binary/multiple systems. This DROD fragmentation scenario is a more dynamic variant of the traditional disk fragmentation scenario for binary/multiple formation, with dense spiral filaments created by inhomogeneous feeding from the highly structured larger-scale sheetlets rather than the need for angular momentum transport, which is dominated by magnetic braking. 
Provided that the local material is sufficiently demagnetized, with a plasma-$\beta$ of 10 or more, collisions between the dense spiraling filaments play a key role in facilitating gravitational collapse and stellar companion formation by pushing the local magnetic Toomre parameter $Q_\mathrm{m}$ below unity. 
This mechanism can naturally produce {\it in situ} misaligned systems on the 100-au scale, often detected with high-resolution Atacama Large Millimeter Array (ALMA) observations. Our simulations also highlight the importance of non-ideal MHD effects, which affect whether fragmentation occurs and, if so, the masses and orbital parameters of the stellar companions formed. 
\end{abstract}



%
%

\section{Introduction}
\label{sec:introduction}

The majority of stars are in binary or multiple systems. How binaries/multiples form is a fundamental problem in star formation with a long and distinguished history, as reviewed by, e.g., \citet{Tohline2002} and \citet{Offner2023}. At the heart of the problem lies how the star-forming gas is condensed and how the condensed gas evolves into independent centers of gravitational collapse, a process commonly referred to as fragmentation. 

Early work focused on the role of the rotation rate and the degree of elongation of dense collapsing cloud cores \citep[e.g.,][]{Tohline2002}. More recent work included turbulence on relatively large scales, which creates independent collapse centers, typically yielding binary/multiple systems of relatively wide separations \citep[e.g.][]{Offner2016, Kuffmeier2019}. It is widely believed that binary/multiple systems on the 100-au scale or closer are formed primarily through disk fragmentation \citep{Offner2023}. However, there is an increasing number of 100-au scale binaries/multiples whose disks are significantly misaligned, e.g., HK Tau \citep{Jensen2014}, L1448 IRAS2A \citep{Tobin2015}, and VLA1623 \citep{Ohashi2022}. It is difficult to form such misaligned systems via the traditional scenario of fragmentation of a well-ordered, co-planar disk. This paper aims to develop a more dynamic picture of the fragmentation of the so-called "Dense ROtation-Dominated structures" (DROD structures or DRODs for short) that are much more inhomogeneous and dynamically active than the traditional disks and that can naturally produce misaligned binary/multiple systems. In this picture, a strong interplay between magnetic fields, turbulence, and gravity drives the inhomogeneities and dynamic activities of the dense fragmenting gas.  

Magnetic fields are expected to play a crucial role in binary/multiple star formation because they can efficiently remove the angular momentum needed for binary/multiple formation through magnetic braking, potentially suppressing fragmentation altogether, leading to the so-called "fragmentation crisis" in the ideal MHD limit \citep{Hennebelle2008}. Similarly, efficient angular momentum removal during core collapse can potentially suppress disk formation, leading to the so-called "magnetic braking catastrophe" \citep{Li2014PPVI}. Both crises highlight the importance of non-ideal MHD effects, which weaken the coupling between the magnetic field and the bulk neutral gas (and thus the magnetic braking) in both disk and binary/multiple formation. Thus, non-ideal MHD effects are included in our investigation. 

Previous studies have stressed different aspects of the role of magnetic fields in binary/multiple formation. For example, \citet{Machida2008} and \citet{Price2008} explored the effects of the initial magnetic field strength on the number of stellar objects formed, and their subsequent evolution, whereas \citet{Harada2021} stressed the importance of the relative orientation between the magnetic field and the rotation axis. Non-ideal MHD effects were explored extensively in the related problem of disk formation (as reviewed in, e.g., \citealp[][]{Li2014PPVI} and \citealp{Tsukamoto2023}), but less so in the context of binary/multiple formation (see, however, \citealp{Wurster2019a} and \citealp{Mignon-Risse2023}). Exploring the role of non-ideal MHD, particularly ambipolar diffusion, is a focus of this investigation. 

Specifically, we will concentrate on how the interplay between the magnetic field, turbulence, and gravity affects binary/multiple star formation in the presence of ambipolar diffusion. We have previously demonstrated that the interplay produces the so-called "gravo-magneto-sheetlets" that dominate the mass and angular momentum contents of the inner protostellar envelope that feeds the central protostellar disk \citep[][see their Fig.3 for a 3D visualization of the structure]{Lam2019, Tu2023}; they are the strongly perturbed version of the classic pseudo-disk of \citet{Galli1993} that has long been recognized to play a central role in disk formation \citep[e.g.,][]{Xu2021}. Given how closely related the problems of disk and binary/multiple star formation are, it is natural to expect the "gravo-magneto-sheetlets" ("sheetlets" for short hereafter) to play an important role in the binary/multiple formation as well. In this paper, we will demonstrate that this is indeed the case. 

In particular, we will show that the highly dynamic collapsing sheetlets feed a highly dynamic dense rotation-dominated structure (DROD) around the primary star on a smaller, 100-au scale, which produces stellar companions through collisions of dense spiraling filaments. The rest of the paper is organized as follows. In \S~\ref{sec:Setup}, we describe the problem setup.  
It is followed by an overview of the simulation results in \S~\ref{sec:overview}. In \S~\ref{sec:DenseGas}, we describe the transformation of the collapsing sheetlets that dominate the structure and dynamics of the inner protostellar envelope into the dense rotation-dominated structures that set the stage for fragmentation. It is followed by a discussion of the conditions for fragmentation and stellar companion formation in \S~\ref{sec:ConditionsForFragmentation}. Our main results and their observational implications are discussed in \S~\ref{sec:discussion} and summarized in \S~\ref{sec:conclusion}.
 
\section{Problem Setup}
\label{sec:Setup}

The formation of stars and their associated structures through the collapse of a molecular cloud core is a complex process involving non-ideal magneto-hydrodynamics (particularly ambipolar diffusion) and turbulence. The following set of equations governs the process:
\begin{equation}
    \frac{\partial\rho}{\partial t} + \nabla\cdot(\rho\mathbfit{v}) = 0,
\end{equation}
\begin{equation}
    \rho\frac{\partial \mathbfit{v}}{\partial t} + \rho(\mathbfit{v}\cdot\nabla)\mathbfit{v} = -\nabla P + \frac{1}{c}\mathbfit{J}\times\mathbfit{B} - \rho\nabla\Phi_g,
    \label{equ:mhd momentum equation}
\end{equation}
\begin{equation}
    \frac{\partial \mathbfit{B}}{\partial t} = \nabla\times(\mathbfit{v}\times\mathbfit{B}) - \frac{4\pi}{c}\nabla\times(\eta_A\mathbfit{J}_\perp),
\end{equation}
\begin{equation}
    \nabla^2\Phi_g = 4\pi G\rho,
\end{equation}
where $\mathbfit{J} = (c/4\pi)\nabla\times\mathbfit{B}$ is the current density, $\mathbfit{J}_\perp = [(\mathbfit{J}\times\mathbfit{B})\times\mathbfit{B}] / B^2$ the component of current density perpendicular to the magnetic field, and $\eta_A$ the ambipolar diffusivity. Other symbols have their usual meanings.  

Our setup closely follows that of \citet{Tu2023} and \citet{Lam2019}. In particular, we use the ATHENA++ code with adaptive mesh refinement (AMR) and the full multigrid (FMG) mode of the multigrid self-gravity solver. The boundary conditions of the self-gravity solver are obtained using multipole expansion for the gravitation potential \citep{Stone2020, Tomida2023}. The simulations are conducted in Cartesian coordinates, with outflow boundary conditions employed at all simulation boundaries.

We use five techniques to speed up the calculations. 
First, we use the same simplified, four power-laws equation of state (EOS) in \citet{Tu2023}. The adiabatic constant as a function of density is divided into four segments, with $\gamma_1=1$, $\gamma_2=1.1$, $\gamma_3=1.4$ and $\gamma_4=5/3$ separated by densities $\rho_{1-2} = 10^{-13}\mathrm{g\ cm^{-3}}$, $\rho_{2-3}=3.16\times10^{-12}\mathrm{g\ cm^{-3}}$ and $\rho_{3-4}=5.66\times10^{-9}\mathrm{g\ cm^{-3}}$ respectively. 
Second, we use the AMR capacity to concentrate computational resources in the densest, dynamically active region. A refinement level is added when the local cell size exceeds 1/16 of the local Jeans length. In the most refined region, the cell size is about 1.22 au.
Third, we use the same sink particle treatment as in \cite{Tu2023} to avoid prohibitively small simulation time-step and represent protostars' formation. A sink particle is created when the simulation reaches the highest refinement level and the local cell size exceeds 1/8 of the local Jeans length. The formed sink particle then accretes both mass and momentum from its nearby $3\times3\times3$ cells. If more than one sink particle is created, two sink particles would merge only if the two sink particles are separated by less than 2 cells (about 2.4 au). 
Fourth, we cap the ambipolar diffusion coefficient using
equation (7) of \citet{Lam2019} to avoid prohibitively small time steps. Finally, we cap the Alfvén speed by adding small amounts of mass in highly evacuated cells. The last two treatments primarily affect very low-density regions, and we monitor the simulation to ensure the mass added stays negligible compared to the mass in the computation domain.

The simulation is initialized by putting a 1.0 $M_\odot$ perturbed pseudo-Bonner-Ebert sphere of radius $r_\mathrm{max} = 2000\ \mathrm{au}$ at the center of the 10,000 au size cubic simulation box. The initial density is prescribed by
\begin{equation}
    \rho(r) = \frac{\rho_0}{1 + (r/r_c)} (1 + \delta_p)
\end{equation}
where
\begin{equation}
    \delta_p = A_\theta \frac{r^2}{r_\mathrm{max}^2} \cos(2\theta)
\end{equation}
where $r$ is the spherical radius, $\theta$ the polar angle from the $z-$axis, and $A_\theta=0.2$ determines the magnitude of the initial density perturbation. We chose a characteristic radius of $r_c = 1,000$ au, which yields a central density of $\rho_0 = 5.29\times10^{-17}\ \mathrm{g\ cm^{-3}}$. The rest of the space is filled with low-density gas of $4.56\times10^{-20}\ \mathrm{g\ cm^{-3}}$. We assume the core has an initial solid body rotation around the $z$-axis, with a rate of $1.27\times10^{-12}\ \mathrm{s}^{-1}$. The setup yields a ratio between the total gas thermal energy and the absolute value of the total gravitational energy of $\alpha = 0.32$ and a ratio between the rotational energy and the absolute value of the total gravitational energy of $\beta_0 = 0.08$. The initial uniform $z$-direction magnetic field has a strength of $3.28\times10^{-4}\ \mathrm{G}$, corresponding to a dimensionless mass-to-flux ratio about 3.5 in units of the critical value $1/(2\pi G^{1/2})$. 

To simulate a turbulent pre-stellar core \citep{Bergin2007}, we include an initial turbulence in our models, as in \citet{Lam2019} and \citet{Tu2023}. The turbulence could be inherited from the larger-scale cloud in which the core is embedded. A root-mean-square Mach number $\mathcal{M}=1$ is obtained by normalizing the velocity field's amplitude for all of the turbulent models. One difference between the models in this paper and those in \citet{Tu2023} is the angular momentum of the turbulence isn't removed from the initial rotation. The angular momentum in the turbulence results in a tilted structure in this paper.

Ambipolar diffusivity is treated similarly as in \citet{Tu2023} to parameterize the uncertainty in the ambipolar diffusivity coefficient associated with the uncertainties in the ionization level (caused by, e.g., uncertain grain size distribution and cosmic ray ionization rate). Following \citet{Lam2019}, we choose to parameterize it in the following way for efficient parameter exploration: 
\begin{equation}
    \eta_A = \frac{B^2}{4\pi\gamma\rho\rho_i}\label{equ:ambipolar_coeff},
\end{equation}
where $\gamma=\langle\sigma v\rangle/(m + m_i)$ is the ion-neutral drag coefficient, and $\rho_i = C\rho^{1/2}$ is the ion density, with the coefficient $C$ proportional to the square root of the cosmic ray ionization rate \citep{Shu1992,Draine1983}. With these approximations, the ambipolar diffusivity can be written as
\begin{equation}
    \eta_A = \eta_0 \frac{B^2}{4\pi\rho^{3/2}},
    \label{equ:etaq}
\end{equation}
where $\eta_0 = \Tilde{\eta}_0/(\gamma C)$, with  $\gamma=3.5\times 10^{13}\mathrm{cm^3g^{-1}s^{-1}}$ and $C=3\times10^{-16}~{\rm cm}^{-3/2}~{\rm g}^{1/2}$ \citep{Shu1992}. The $\Tilde{\eta}_0$ is a free scaling parameter for the ambipolar diffusivity coefficient, with $\Tilde{\eta}_0=1$ corresponding to the standard cosmic ionization rate of $10^{-17}$~s$^{-1}$. 
We consider three values of $\Tilde{\eta}_0$: $1$, $2$, and $10$ (see Table~\ref{tab:para}). In addition, we consider a variant of the $\Tilde{\eta}_0=2$ model that does not have any initial turbulence (Model M0.0AD2.0) and another variant where the turbulence Mach number is twice as large and the initial magnetic field and rotation axis are perpendicular (Model PRM2.0AD2.0). 

\begin{table}
    \centering
    \begin{tabular}{c|c|c|c|c|c|l}
        Model & $\mathcal{M}$ & $\mathrm{\Tilde{\eta}_0}$ & $N_\mathrm{star}$ & $N_\mathrm{tot}$ & $t_\mathrm{primary}$ & comment \\
        \hline
        M1.0AD2.0 & 1 & 2 & 5 & 6 & 13560 & reference \\
        M0.0AD2.0 & 0 & 2 & 3 & 3 & 12050 & no turbulence \\
        M1.0AD1.0 & 1 & 1 & 1 & 1 & 13670 & low diffusivity \\
        M1.0AD10  & 1 & 10 & 5 & 6 & 13110 & high diffusivity \\
        PRM2.0AD2.0 & 2 & 2 & 2 & 2 & 16290 & $\mathbf{\Omega}_0\perp \mathbfit{B}_0$
    \end{tabular}
    \caption{Summary of each simulation model's initial conditions and outcome. $\mathcal{M}$ is the Mach number of the initial turbulence; $\Tilde{\eta}_0$ is the scaling of the ambipolar diffusivity coefficient defined in equ.~\ref{equ:etaq}; $N_\mathrm{star}$ is the number of stars at the end of the simulation (excluding merged stars; stellar companion 2 merged with stellar companion 1 in Model M1.0AD2.0, and stellar companion 5 merged with stellar companion 4 in Model M1.0AD10); $N_\mathrm{tot}$ is the total number of stars formed in the simulation (including merged stars); $t_\mathrm{primary}$ is the formation time of the primary in years.}
    \label{tab:para}
\end{table}
 
\subsection{Post-process Lagrangian tracer particles}
\label{sec:PostProcessing}

To better understand the gas evolution during the simulation, particularly in the period leading up to stellar companion formation (see fig.~\ref{fig:Collision_3D} below), we developed a post-process Lagrangian tracer particle code to trace the gas trajectory. 
%
We solve the equation of motion of tracer particles and update their velocities and positions at every time step. The equation of motion is given by
\begin{equation}
    \frac{d\mathbfit{x}}{dt} = \mathbfit{v}_\mathrm{gas}
\end{equation}
where $\mathbfit{v}_\mathrm{gas}$ is the instantaneous gas velocity at the location of the tracer particle. We obtain the gas property information at every location ($\mathbfit{x}$) at every time step ($t$) in two steps.  Firstly, we interpolate the gas properties spatially using the Triangular-Shaped-Cloud (TSC) interpolation method \citep{TSC} at the hydro frames immediately before (denoted with a subscript "1", e.g., $v_1$ and $t_1$) and after (denoted with a subscript "2", e.g., $v_2$ and $t_2$) the time of interest. Secondly, we get the hydro quantities at the time of interest using a linear interpolation of their values at $t_1$ and $t_2$, e.g.,
\begin{equation}
    v_t = v_1 + \frac{(v_2 - v_1)(t - t_1)}{t_2 - t_1}
\end{equation}
Finally, we solve the equation of motion using the second-order Runge-Kutta method.

To maximize the efficiency of the post-process simulation, we allow each particle to evolve using its own time steps by taking advantage of the fact that the hydro information is known. We only require the time steps for all particles to synchronize at the time of output. The time step for each particle is determined by its instantaneous speed $\mathbfit{v}_t$ and the local cell size $dx_\mathrm{grid}$, and is given by
\begin{equation}
    dt = 0.3\times\min\Big(dt_\mathrm{max}, \frac{dx_\mathrm{grid}}{|\mathbfit{v}_t|}\Big)
\end{equation}
This guarantees that a particle travels no more than 0.3 times the local cell size each time step \citep[similar to the CFL condition described in][]{CFL1928}. The maximum time step $dt_\mathrm{max} = 1\mathrm{\ yr}$ caps the time step so that it does not become too large when the tracer is almost stationary. 

\section{Overview of Results}
\label{sec:overview}

\begin{figure*}
    \centering
    \includegraphics[width=\textwidth]{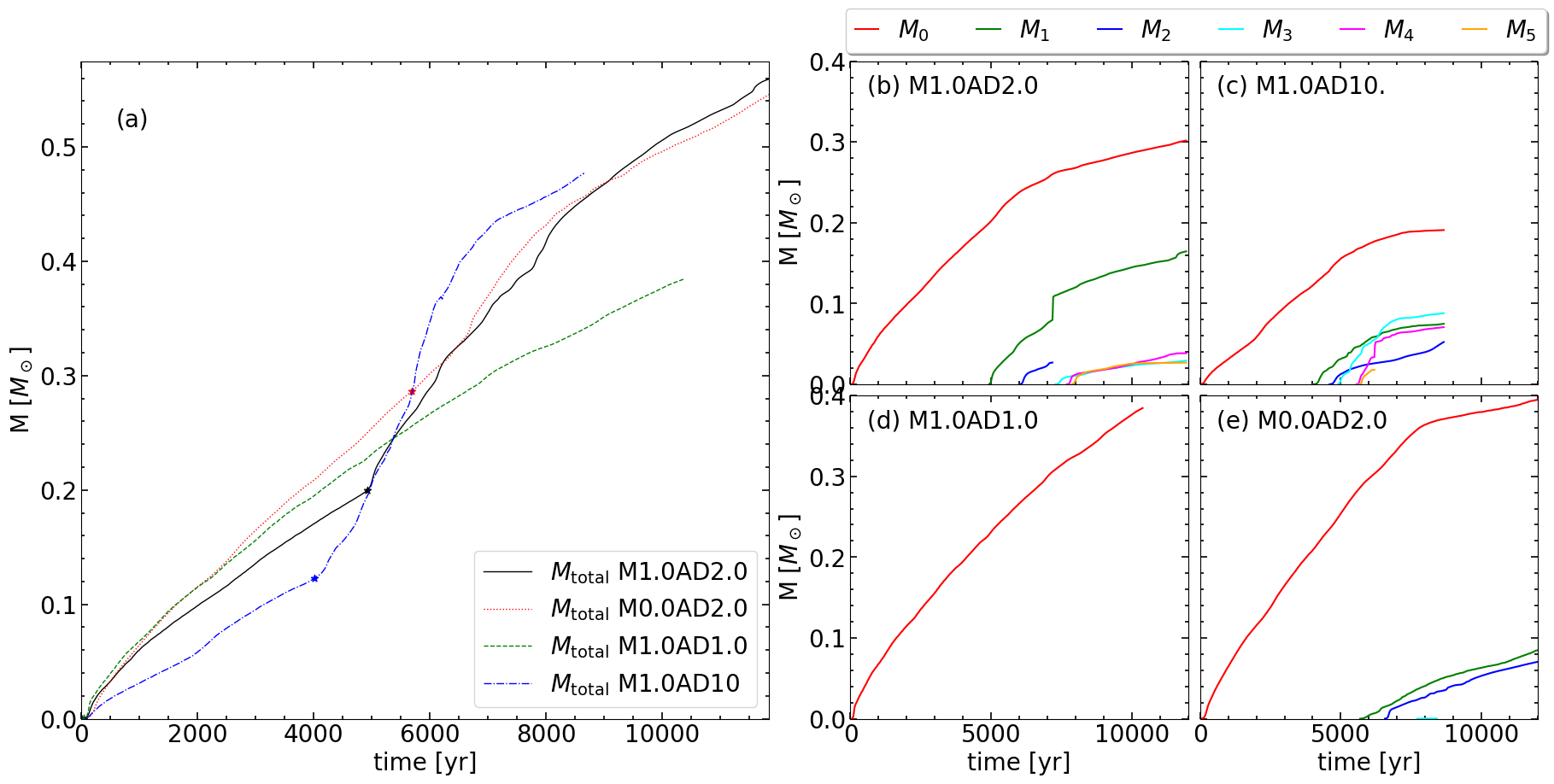}
    \caption{\textbf{Left Panel (a): } The time evolution of the total mass of all stars formed in each of the first four models listed in Table~\ref{tab:para}, starting from the time of the primary formation. The "star" on each curve marks the formation of the secondary, if any. \textbf{Right Panels (b)-(e):} The time evolution of the masses of individual stars for the four models. The quantity $M_0$ above the panels refers to the mass of the primary, $M_1$ that of the first stellar companion (or the secondary), and $M_2$ that of the second stellar companion, etc.} 
    \label{fig:mass_summary}
\end{figure*}

\begin{figure*}
    \centering
   \includegraphics[width=\textwidth]{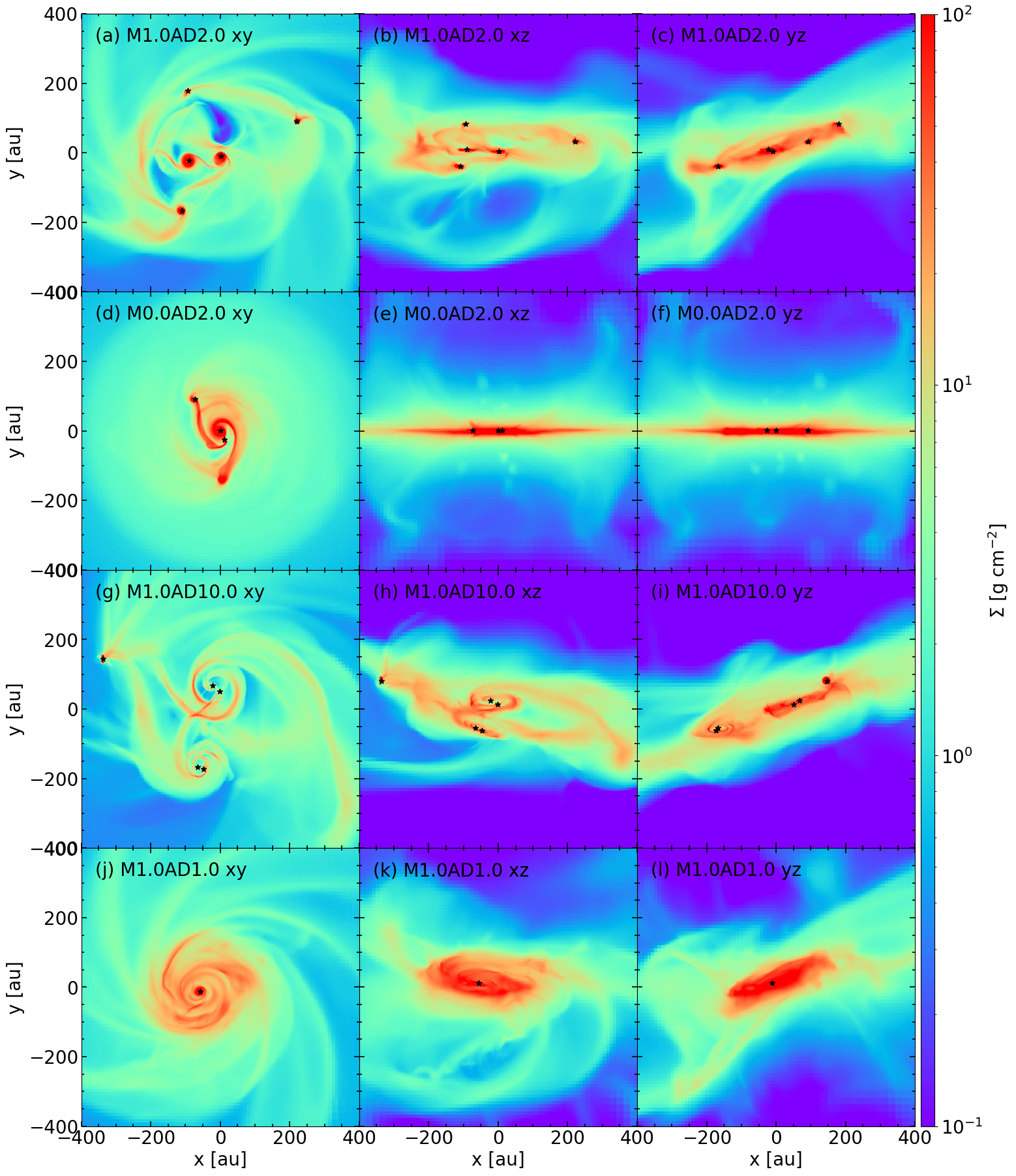}
    \caption{Column density distributions within a $800\mathrm{\ au}^3$ central region along the $\hat{z}-$ (left column), $\hat{y}-$ (middle column), and $\hat{x}-$direction (right column), respectively at a representative time, with the model name shown in each panel. The ``stars'' in each panel mark the projected locations of the sink particles. An animation of this figure can be found at \url{https://figshare.com/s/e96a9f3b77e496d4e814}.}
    \label{fig:prime_plane_slice}
\end{figure*}

In this section, we give a general impression of the simulation results by highlighting some broad features in Fig.~\ref{fig:mass_summary} and \ref{fig:prime_plane_slice}. Fig.~\ref{fig:mass_summary} shows the time evolution of the total mass of all protostellar objects (sinks) after the formation of the first object (referred to as the ``primary" hereafter) in each of the first four models listed in Table~\ref{tab:para} (left panel) and the mass evolution of each individual object (right panels); the last model in the table (PRM2.0AD2.0) will be discussed separately in \S~\ref{sec:discussion} below. The ``star'' on each curve in panel (a)  corresponds to the formation time of the first stellar companion (referred to as the ``secondary" hereafter). In the least magnetically diffusive model (M1.0AD1.0 model), there is no ``star'' on the curve, which means there is only a single object in this model at all times. Only in the more diffusive cases does a secondary form. After the primary formation, the secondary forms the earliest in the M1.0AD10 model and the latest in Model M0.0AD2.0, with Model M1.0AD2.0 in between the two limits. After the secondary formation, the slope of the total stellar mass as a function of time changes suddenly because mass is now accreted onto the second star in addition to the primary (Fig.~\ref{fig:mass_summary}a). Because Model M1.0AD2.0 has multiple stellar objects formed at well-defined times (see Fig.~\ref{fig:mass_summary}b), it will be the focus of our detailed analysis and be referred to as the ``reference" model hereafter. 

To give a general impression of the spatial distribution of the protostars and the gas structures around them, we show in Fig.~\ref{fig:prime_plane_slice} the column density distribution and the projected sink particle positions at a representative time when all stars have formed and grown beyond 0.02$M_\odot$. Each plot is centered on the center-of-mass of the simulation. Fig.~\ref{fig:prime_plane_slice} shows that most of the dense gas and formed stars reside in a relatively flat structure, which is particularly evident when viewed along the $\hat{x}-$direction (see the right panels). In the simplest case of the non-turbulent model M0.0AD2.0, the dense flattened structure lies on the equatorial plane perpendicular to the initial magnetic field and rotation axis (see Fig.~\ref{fig:prime_plane_slice}e,f). It is tilted and modified by the turbulence included in the reference model M1.0AD2.0, which apparently facilitated an earlier formation of the secondary (see Fig.~\ref{fig:mass_summary}a) and the formation of more companions at later times (Fig.~\ref{fig:mass_summary}b). The comparison of these two models clearly illustrates the beneficial effects of turbulence on the fragmentation of dense gas that leads to the formation of binary/multiple systems. 
%
%

The dense gas fragmentation also depends on the magnetic diffusivity. The dependence is illustrated most vividly by the least magnetic diffusive model M1.0AD1.0, where fragmentation does not occur at all, despite the presence of initial turbulence as in the reference model. In contrast, several stars are formed in the more magnetically diffusive models, Models M1.0AD2.0 and M1.0AD10. This result is consistent with the finding of earlier work that a dynamically important magnetic field tends to suppress fragmentation in the ideal MHD limit, leading to a potential binary/multiple star formation crisis \citep[e.g.][]{Hennebelle2008, Boss2013, Zhao2013}. Our models demonstrate that non-ideal MHD effects (specifically ambipolar diffusion) can avert the fragmentation crisis, provided the magnetic diffusivity is large enough. The value of the diffusivity can also strongly affect the properties of the formed stellar systems (such as the number of stars and their mass distribution and orbital parameters) when fragmentation does occur, as can be seen from a comparison of Model M1.0AD2.0 (top row of Fig.~\ref{fig:prime_plane_slice} and Fig.~\ref{fig:mass_summary}b) and Model M1.0AD10 (third row of Fig.~\ref{fig:prime_plane_slice} and Fig.~\ref{fig:mass_summary}c), whose diffusivities differ by a factor of 5. 

Since the formation and evolution of dense structures lie at the heart of gas fragmentation and binary/multiple formation, they will be discussed first (\S~\ref{sec:DenseGas}). It is followed by a discussion of the conditions for binary/multiple formation (\S~\ref{sec:ConditionsForFragmentation}). 

\begin{figure*}
   \includegraphics[width=\textwidth]{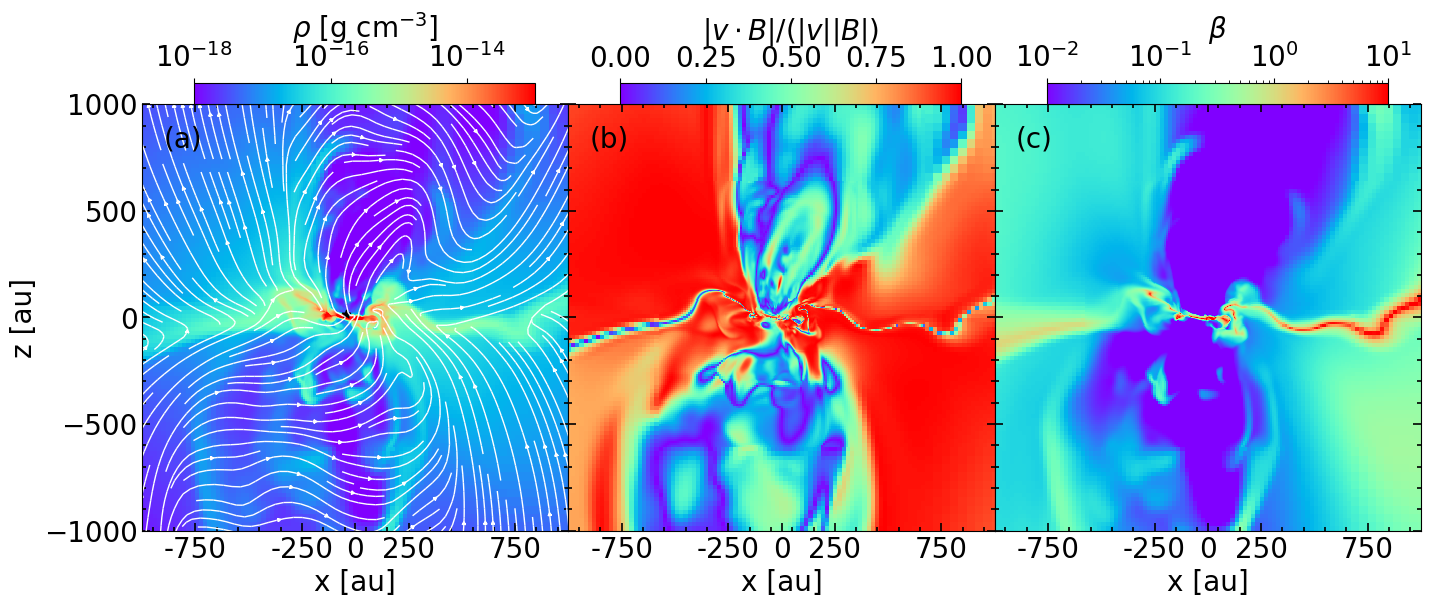}
    \caption{Dense demagnetized sheetlets embedded in a strongly magnetized diffuse background. Plotted are the distributions of the mass density (panel [a], with field lines superposed), the quantity $\cos(\theta_\mathrm{ali})$ (panel [b], where $\theta_\mathrm{ali}$ is the angle between the magnetic and velocity field), and the plasma-$\beta$ (panel [c]) on an $xz$-slice through the reference model at a representative time on a relatively large ($\sim$1000~au) scale. The thin, wiggly, bluish filaments in the middle panel (where dense demagnetized gas collapses across field lines) mark the 2D ($xz$-)cross-section of the 3D gravo-magneto-sheetlets. The low-density, strongly magnetically dominated cavities above and below the sheetlets are magnetically driven outflow lobes. }
    \label{fig:denest_slices}
\end{figure*}

\begin{figure*}
 \centering
  \begin{minipage}{0.48\textwidth}
     \includegraphics[width=\textwidth]{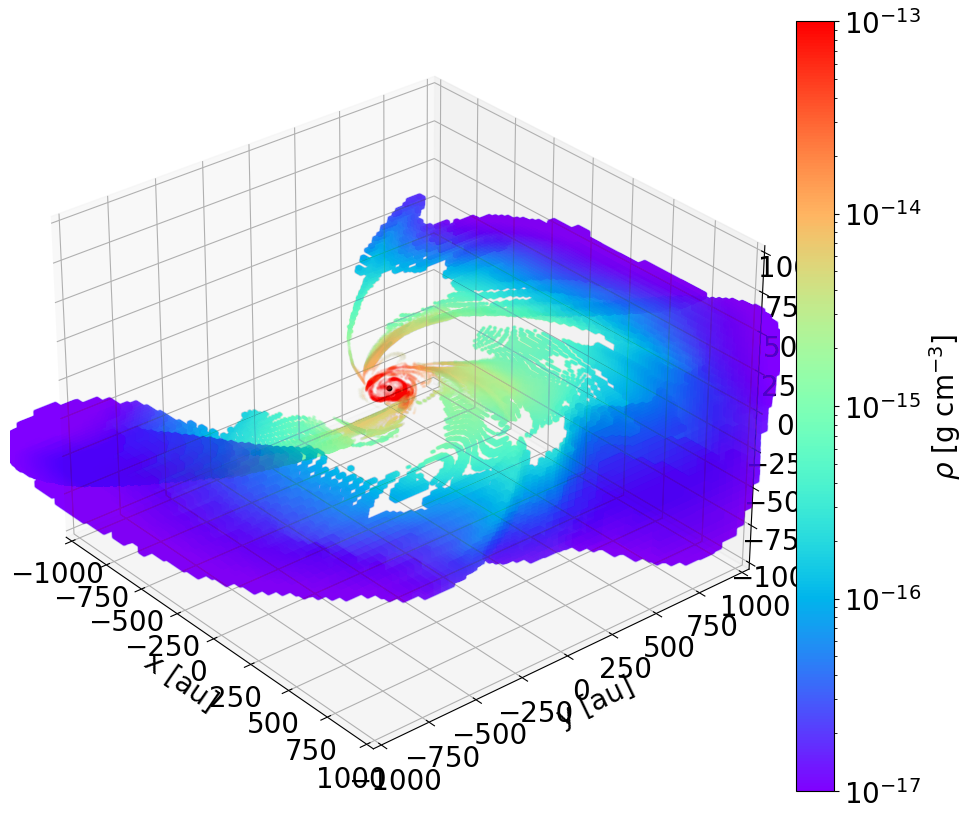}
    \end{minipage}
    \begin{minipage}{0.48\textwidth}
       \includegraphics[width=\textwidth]{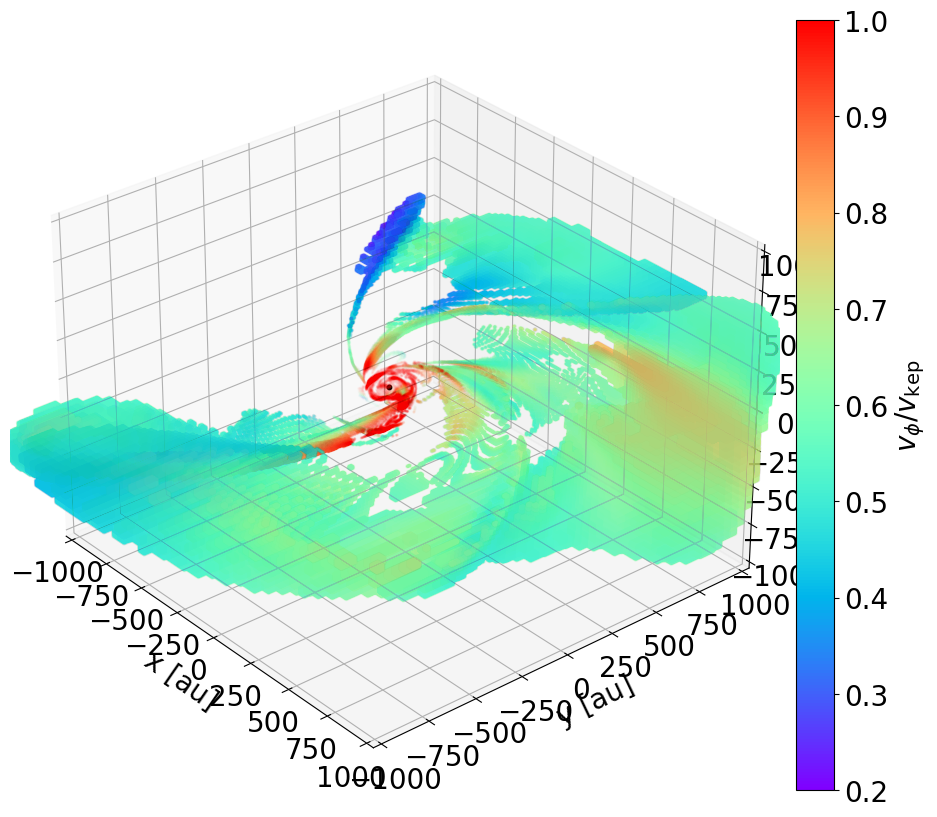}
    \end{minipage}
    \caption{Sheetlets in 3D. Plotted are the 3D structures of the gravo-magneto-sheetlets at the representative time for the reference model shown in Fig.~\ref{fig:denest_slices}. The color in panel (a) shows the local density. It shows the rotation speed around the center of the mass normalized by the local Keplerian speed in panel (b). Note the large-scale, relatively open spirals in the flattened sheetlets, which turn into more tightly wound spirals on the smaller scale. An animation of the figure at different times can be found at \url{https://figshare.com/s/e9824528e3a453be5645} (for the left panel) and \url{https://figshare.com/s/cb4a7d3e412d5972eb57} (for the right panel). An animation of the left panel viewed from different angles can be found at \url{https://figshare.com/s/68459e6b519c382596e8}. 
    }
    \label{fig:sheetlet_3D}
\end{figure*}

\section{Origins of Dense Binary/Multiple-Forming Gas Structures}
\label{sec:DenseGas}

\subsection{Large-Scale Collapsing Gravo-magneto-Sheetlets}
\label{sec:sheetlets}

The concentration of gas in a collapsing molecular cloud core and the subsequent evolution of the dense gas set the stage for fragmentation and binary/multiple formation. Both processes are shaped by a complex interplay between several processes, including gravitational collapse, rotation, turbulence, and the magnetic field. In particular, in the presence of an ordered, dynamically important magnetic field, the gravitational collapse tends to funnel the collapsing material along the field lines towards a dense flattened structure that is spatially coherent - the classic ``pseudodisk" in the simplest case of the collapse of a non-turbulent initially uniformly magnetized singular isothermal sphere \citep{Galli1993}. This product of the interplay between the gravitational collapse and anisotropic magnetic resistance to it shows up clearly in our non-turbulent model M0.0AD2.0 as a dense equatorial sheet that is the main conduit for delivering the dense gas close to the center, where fragmentation leading to binary/multiple systems occurs (see Fig.~\ref{fig:prime_plane_slice}, second row). In the presence of turbulence, the flat gravo-magneto-sheet is warped and fragmented into more irregular gravo-magneto-sheetlets, which show up in the column density maps as filaments connecting to the densest region near the center (see, e.g., Fig.~\ref{fig:prime_plane_slice}c,i). 
\citet{Tu2023} showed that these filamentary column-density structures are spatially connected in 3D, which they termed ``gravo-magneto-sheetlets" (or "sheetlets" for short; see their Fig.~3 for a 3D visualization of the structure). 

\citet{Tu2023} found two defining characteristics of the sheetlets: (1) they are less magnetized compared to the background (with a significantly higher plasma-$\beta$) and (2) they tend to move across the field lines, with a relatively small value of the quantity $\cos\theta_\mathrm{ali}=|\mathbfit{v}\cdot\mathbfit{B}|/(|\mathbfit{v}||\mathbfit{B}|)$, where $\theta_\mathrm{ali}$ is the angle between the magnetic field and velocity field. The latter contrasts with the lower-density background, where the strongly magnetized (low plasma-$\beta$) gas tends to flow along the field lines, with $\cos\theta_\mathrm{ali}$ close to unity. 
These two characteristics are illustrated in Fig.~\ref{fig:denest_slices}, which shows the distributions of the mass density, $\cos\theta_\mathrm{ali}$, and the plasma-$\beta$ of the reference model M1.0AD2.0 on a $xz$-slide through the primary at a representative time $t=19,000$~years after the start of the simulation, when the primary mass reached $M_0=0.2M_\odot$.  
The most striking feature is present in the distribution of $\cos\theta_\mathrm{ali}$ in the middle panels where thin wiggly blue filaments of small $\cos\theta_\mathrm{ali}$ are embedded in a red background of $\cos\theta_\mathrm{ali}\approx 1$. These filaments have higher plasma-$\beta$ (see right panels) and higher density (left panels) than the surrounding background. They also have sharp pinching of the local field lines (see the white lines in the left panels), which is produced by the field dragging by the cross-field gravitational collapse, as expected in the dense sheetlets \citep{Tu2023}\footnote{We note dense filaments similar to those shown in Fig.~\ref{fig:prime_plane_slice} are also observed in disk formation simulations starting from larger cloud scales that include magnetic fields and turbulence \citep[e.g.,][]{Kuffmeier2017,Lebreuilly2021}. At least some of them are likely gravo-magneto-sheetlets, but we do not have detailed information on their magnetic and velocity fields to be sure.}.

To show the true shape of the turbulence-perturbed gravo-magneto-sheetlets in 3D, we make use of their two defining characteristics and search for all cells where the plasma-$\beta$ > 2.51 and $\cos\theta_\mathrm{ali} < 0.6$. Fig.~\ref{fig:sheetlet_3D} displays a 3D visualization of the ridges of the sheetlets identified with the above two criteria in the reference model at the same time ($t=19,000$~years, corresponding to 5500 years after the primary formation) as shown in Fig.~\ref{fig:denest_slices} on a relatively large scale of $1000$~au. Note that with these two selection criteria, we are picking out only the densest part (the ``spline'') of the sheetlet identified in \citet{Tu2023}, where the sheetlet is identified by a density threshold using mass contained. The reader is encouraged to view the animation of the figure from different viewing angles to get a more vivid impression of this crucial 3D structure, which is the main conduit to funnel the dense material collected from the more diffuse regions of the protostellar envelope along magnetic field lines to the central region where fragmentation and binary/multiple star formation occurs {\citep[e.g.,][their Fig.5e]{Tu2023}}. It contains several large-scale spirals with relatively open arms, which are caused by a relatively fast gravitational collapse compared to the rotation. The gravitational collapse is integral to the density increase in the sheetlets. However, it limits the mass the sheetlets can collect to form binary/multiple systems locally through gravitational instability because of mass continuity: for a given mass accretion rate, the faster the infall is, the lower the local density becomes. To collect enough mass to form binary/multiple systems, another mechanism must come into play to slow down the collapse and force the matter to pile up further. It is, of course, through rotational support against gravity, as we demonstrate next. 

\subsection{Dense Rotation-Dominated Structures}
\label{sec:RotationPileUp}

The transformation of the large-scale dense collapsing gravo-magneto-sheetlets into even denser, smaller-scale rotation-dominated structures is illustrated in the right panel of Fig.~\ref{fig:sheetlet_3D}, which shows (in color) the non-radial speed around the primary star normalized by the local Keplerian speed. Because the primary's mass is relatively small compared to that of the inner envelope at early times, we approximate the local Keplerian speed by $v_\mathrm{kep}(r) = \sqrt{GM_\mathrm{in}(r)/r}$ where $r$ is the distance from the primary and $M_\mathrm{in}(r)$ is the mass enclosed within a sphere of radius $r$ (including the stellar mass)\footnote{We caution the reader that this approximation of the local Keplerian velocity assumes a gravitational potential dominated by the first (monopole) term in the multipole expansion, which is more accurate at later times when more mass is concentrated at small radii (including the star/sink).}. A comparison of the left and right panels of Fig.~\ref{fig:sheetlet_3D} clearly shows that the transition from an infall-dominated velocity field to one dominated by rotation corresponds to a jump in density. This rotation-induced mass pileup sets the stage for fragmentation and binary/multiple formation.

\begin{figure}
    \centering
    \includegraphics[width=\columnwidth]{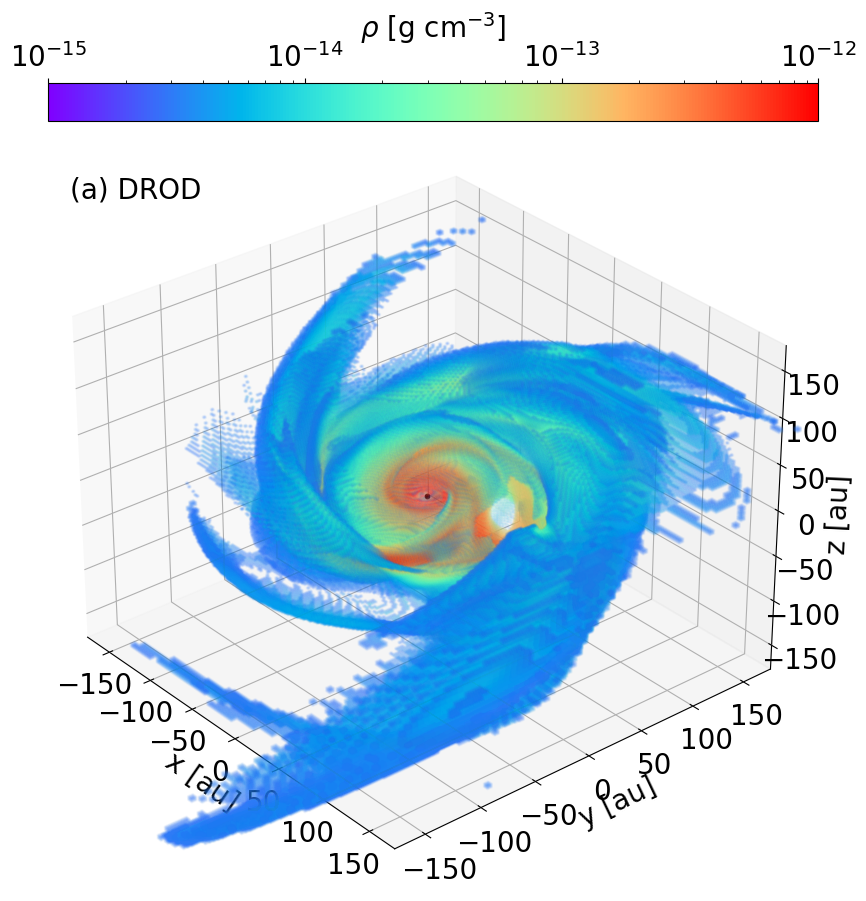}
    \caption{A 3D view of the dense rotation-dominated DROD structure (DROD for short) that is fed by the larger-scale sheetlets; it is the formation site for (often misaligned) stellar companions (see Figs.~\ref{fig:disk_misalignment} and \S~\ref{sec:misalignment} below). 
    An animated version of this figure viewed from different angles can be found at \url{https://figshare.com/s/ff55b7952ead2ba91508}}
    \label{fig:DRODs3D}
\end{figure}

\begin{figure}
    \centering
    \includegraphics[width=\columnwidth]{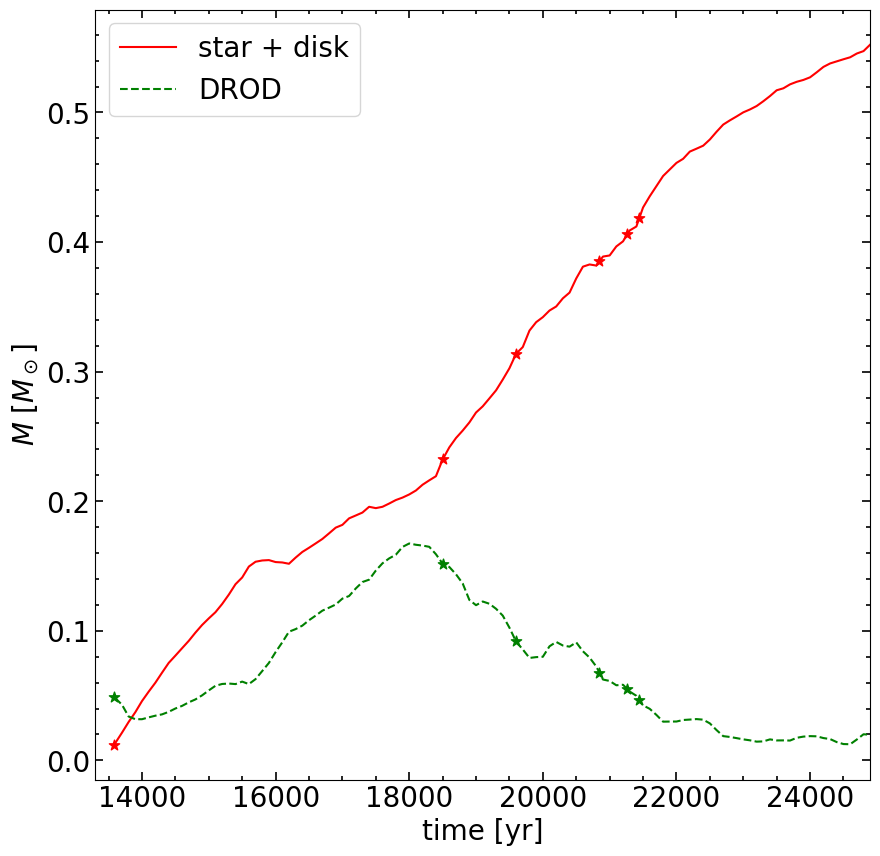}
    \caption{Total mass of the stars and their disks (defined in sec.~\ref{sec:RotationPileUp}, red solid line) and total mass of the DROD (green dashed line). The stars on both curves mark the formation time of each star.}
    \label{fig:diskDRODmass}
\end{figure}

In Fig.~\ref{fig:DRODs3D}, we show the dense rotation-dominated structure with a density greater than $3\times10^{-15}\mathrm{g\ cm^{-3}}$ and rotating at $>90\%$ the local Keplerian speed. We term this structure the "DROD (Dense ROtation-Dominated) structure" (or "DROD" for short) to distinguish it from the traditional circumstellar disk because it is much more irregular and dynamically active (see the online animated version of Fig.~\ref{fig:DRODs3D}). To distinguish between the circumstellar disk and the DROD structure, we define the circumstellar disk to be the gas in the vicinity of a star, rotating at $>90\%$ the local Keplerian speed relative to the host star, and with a density greater than $10^{-12}\ \mathrm{g\ cm^{-3}}$. In Fig.~\ref{fig:diskDRODmass}, we show the mass of the DROD and the star-disk system. The formation times of the stars are marked by the ``star'' on each line.  After the primary formation, the mass of the DROD increases as mass piles up due to rotation. At $\sim 18500$ yr, the DROD starts to fragment into companion stars and 5 additional stellar objects form within the next 3000 yrs. During this rapid star formation era, the mass of the DROD decreases. After the mass becomes depleted in the DROD at $\sim 21500$ yr, no more stars form. This picture could be modified if there is additional material accreting onto the DROD from, e.g., the larger cloud scale \citep[e.g.,][]{Kuffmeier2017,Lebreuilly2021}.
 
Whether the piled-up dense gas in the DROD fragments into a binary/multiple system or not depends on its detailed properties, such as mass and size, which, in the set of models considered in this paper, depend on the magnetic diffusivity. Specifically, the DROD formed in the least magnetically diffusive model, M1.0AD1.0, contains several spirals that connect smoothly to the larger-scale collapsing sheetlets (see Fig.~\ref{fig:prime_plane_slice}j and associated animation). It does not accumulate enough mass to become locally self-gravitating for a long enough period to collapse into companion stellar objects. In contrast, the DROD in each of the three more magnetically diffusive models (M1.0AD2.0, M0.0AD2.0, and M1.0AD10) contains denser, more prominent spiraling filaments that are often spatially connected to the larger-scale collapsing gravo-magneto-sheetlets (see, e.g., Fig.~\ref{fig:prime_plane_slice}a and Fig.~\ref{fig:sheetlet_3D}a and their associated animations) independent of the level of turbulence. Portions of the filaments accumulate enough mass at large enough distances from the primary to collapse locally to form companion stellar objects. 

The above difference can be traced to the effects of magnetic diffusivity on the field strength and configuration and their associated magnetic braking efficiency. These effects have been discussed in depth in \citet{Tu2023} in the context of disk formation around a single star \citep[see also, e.g. ][]{Lam2019, Hennebelle2016, Hennebelle2020, Lee2021}. Specifically, they showed that, as the magnetic diffusivity increases, the magnetic braking becomes less efficient (see their Fig.~12), resulting in a larger (their Fig.~11, top panels) and more massive (their Fig.~1, right panel) disk, which is closer to being gravitationally unstable as measured by the value of the standard Toomre $Q$ parameter (their Fig.~11, bottom panels). This trend makes physical sense since the braking would go to zero in the limit of infinite magnetic diffusivity (i.e., the field is not coupled to the matter at all). The next section shows that it also holds for the DRODs formed in our simulations. 

\section{Conditions for DROD Fragmentation}
\label{sec:ConditionsForFragmentation}

\subsection{Toomre-Q unstable structures}
\label{sec:Q}

A traditional measure for disk fragmentation is the Toomre $Q$ parameter, defined as 
\begin{equation}
    Q = \frac{c_s\Omega}{\pi G \Sigma}
    \label{eq:Q}
\end{equation}
where $c_s$ and $\Sigma$ are the isothermal sound speed and gas column density, respectively, and $\Omega$ is the disk rotation frequency around the central star. Even though the stability condition on the Toomre-$Q$ parameter was initially derived mathematically for axisymmetic perturbations \citep{Toomre1964},  it has often been used to diagnose and predict gravitational fragmentation. Although the DRODs formed in our simulations are not necessarily smooth, well-defined disks (see Fig.~\ref{fig:DRODs3D} and discussion in \S~\ref{sec:nonDiskFrag} below), we will stick with this quantity because it continues to provide a measure of how self-gravitating a region is locally against thermal pressure support and shearing by tidal forces. 

\begin{figure*}
    \centering
    \includegraphics[width=\textwidth]{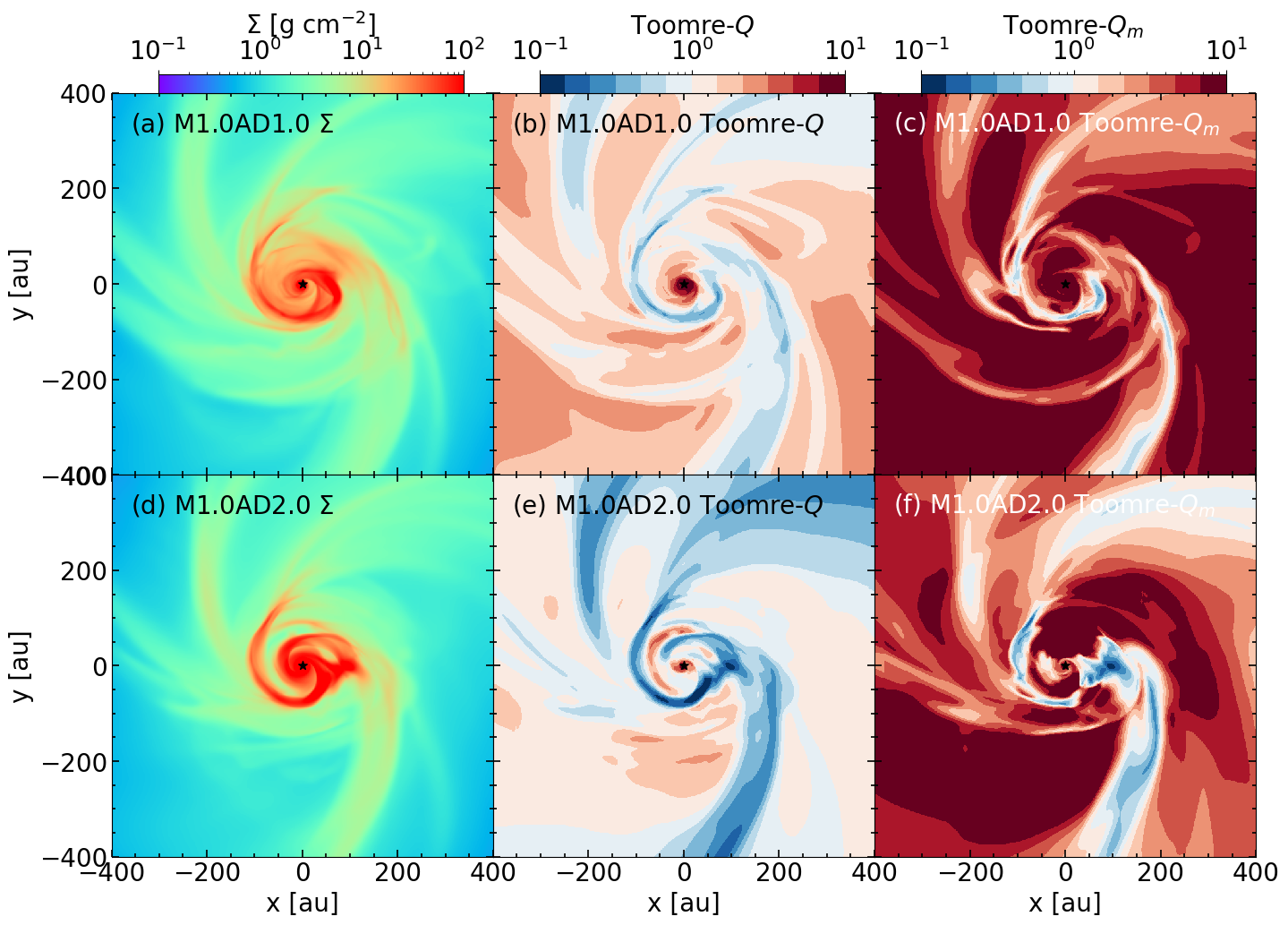}
    \caption{Potential for fragmentation as measured by the Toomre parameter $Q$ (middle panels) and the effects of the magnetic field as measured by the magnetic Toomre parameter $Q_m$ (right panels), for the least diffusive model M1.0AD1.0 (top row) and reference model M1.0AD2.0 (bottom row). The column density is shown in the left panels for comparison. }
    \label{fig:Q}
\end{figure*}

To compute the Toomre $Q$ parameter, we must first determine a ``midplane'' of the simulation, which is the plane that best captures the rotation-dominated structure in each model and is not aligned with any of the coordinate directions. In the least diffusive M1.0AD1.0 model, where only a single star forms, the midplane is defined as the plane through the star that minimizes its distance to the densest 1000 cells, following the definition used in \citet{Tu2023}. To best capture the DRODs in models with multiple stars, we define the midplane as the plane that minimizes its distance to all stars' orbit over all times and goes through the primary star at the time of interest. The sound speed $c_s$ and Keplerian rotation speed $\Omega$ are computed on the midplane. The column density $\Sigma$ is computed by integrating 400~au in both directions perpendicular to the midplane. For the least magnetically diffusive model M1.0AD1.0, we find values of $Q$ well below unity in the prominent spirals in the DROD. The apparently gravitationally unstable spirals often connect smoothly to the larger-scale sheetlets with low values of $Q$, as illustrated in the top middle panel of Fig.~\ref{fig:Q}. At the face value, one may expect the spirals to fragment. However, the DROD remains significantly magnetized, which hinders fragmentation. 

To account for the magnetic effects, we define a magnetic Toomre parameter \citep[following][]{Kim2001}:
\begin{equation}
    Q_m = \frac{\sqrt{c_s^2+v_A^2}\Omega}{\pi G \Sigma}
    \label{eq:Q_m}
\end{equation}
where $v_A$ is the Alfv\'en speed computed on the midplane. It is included to account for approximately the magnetic resistance to local gravitational collapse. 

The magnetic Toomre parameter $Q_m$ in Model M1.0AD1.0 is significantly higher than $Q$, as seen by comparing the top right panel of Fig.~\ref{fig:Q} to the top middle panel. The magnetic effects bring most of the apparently Toomre unstable regions with $Q$ well below unity to marginal stability with $Q_m\sim 1$. In the more diffusive reference model M1.0AD2.0, the apparently Q-unstable regions are more widespread than in Model M1.0AD1.0 (compare the middle panels of Fig.~\ref{fig:Q}). Some of these regions remain unstable even after the magnetic effects are considered. A subset of such apparently unstable regions collapses gravitationally to become stellar companions while others are quickly sheared apart. In the latter case, the relatively strong magnetic field remaining in the DROD plays a key role in suppressing the fragmentation, as discussed further in sec.~\ref{sec:condition-pbeta} below.

The presence of regions with $Q_m < 1$ that do not collapse into stars suggests that the magnetic Toomre criterion is a necessary but not sufficient condition for fragmentation. This conclusion is consistent with the finding of \citet{Mignon-Risse2021} that the Toomre parameter alone cannot determine whether a rotationally supported structure would fragment or not. The following subsection will investigate whether collisions between dense filaments inside the DROD can help induce Toomre-unstable regions to collapse and form stellar companions. 

\subsection{Collision-Induced Fragmentation} 
\label{sec:collisons}


\begin{figure*}
    \centering
    \includegraphics[width=\textwidth]{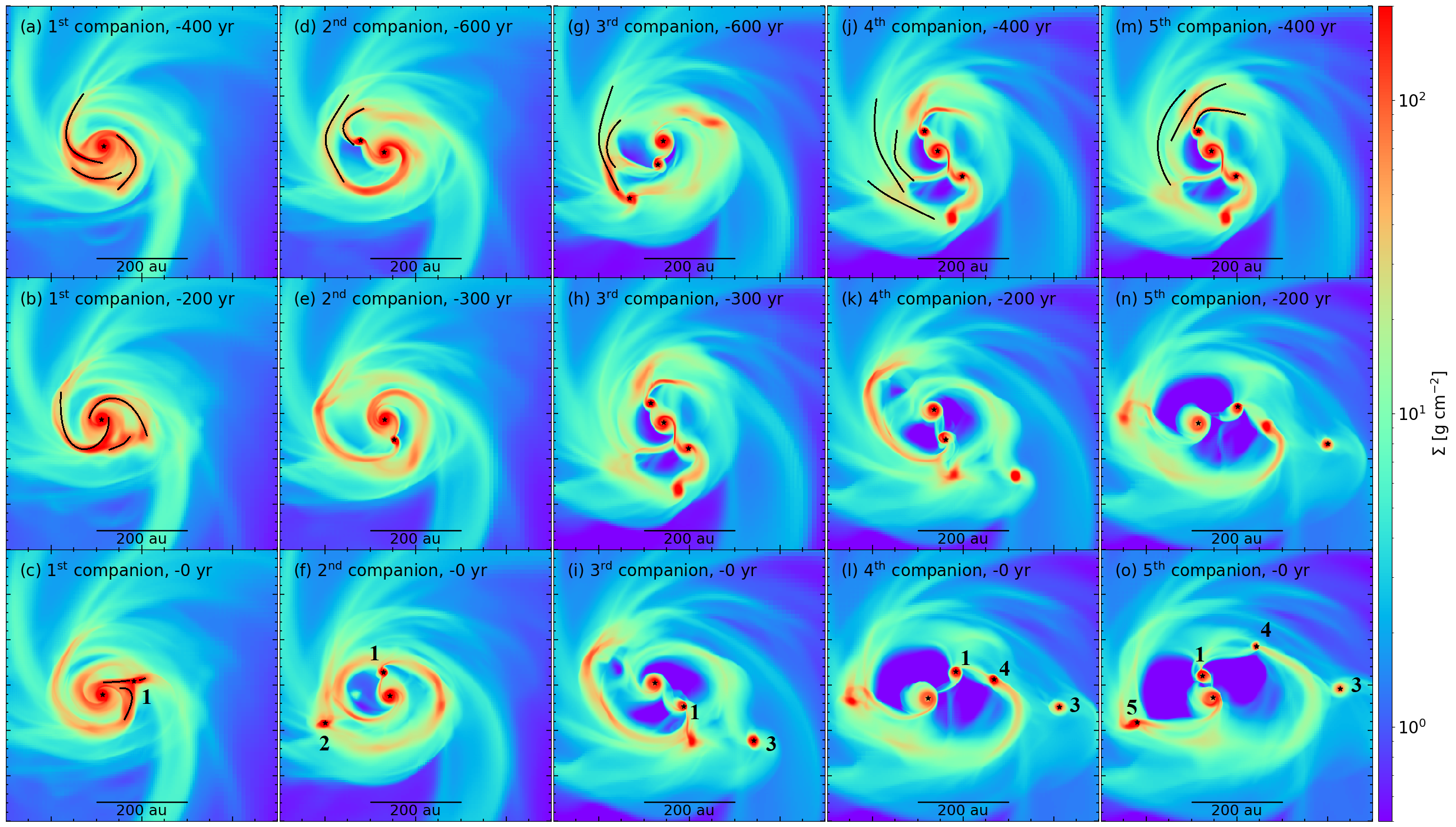}
    \caption{
    Collisions of dense spiraling filaments leading to stellar companion formation. Each column shows three representative snapshots during the formation of each of the five stellar companions of the reference model, highlighting the dense spiraling filaments that collided to form the companion (see the thin black lines). The time before the companion formation is marked in each panel (e.g., -200 yrs). The companions are numbered in the bottom panels. An animation of this figure can be found at \url{https://figshare.com/s/7ff0b55985a7d4336ef2}.
    }
    \label{fig:Collision}
\end{figure*}

The rotation-dominated DRODs formed in our simulations are highly dynamic structures, with dense spiraling filaments often colliding with one another. The collisions are key to forming stellar companions. These collisions are seen most clearly in an animated version of the column density distribution of the DROD along the $z-$axis. A link to the animation can be found in the caption of Fig.~\ref{fig:Collision}, where we highlight the filaments involved in the collisions that form each of the five stellar companions in the reference model, with one column of three panels in the figure for each companion. The formation of the first companion involves the collisions between three dense spiral filaments marked with black line segments in panels (a) through (c). The filament in the middle is compressed by the collision and becomes the first companion. The formation of the second and third companions involves the collision of two dense spiraling filaments and the continued accretion of material along the merged filament after the collision (see the second and third columns, respectively); the second companion later merged with the first companion. Similar to the formation of the first companion, the formation of the fourth and fifth companions each involves three dense spiraling filaments (see the fourth and fifth columns). In the case of the fourth companion, the outer two filaments collide first, with the merged filament subsequently colliding with the innermost filament to form the companion. In the case of the fifth companion, the inner two filaments collide first, with the merged filament colliding with the outermost filament to form the companion. Some low-density cavities appear in the DROD, particularly at later times (see, e.g., the right 3 columns of Fig.~\ref{fig:Collision}). They are likely primarily caused by stellar interactions, as in \cite{Artymowicz1994}, although magnetic fields may have also contributed because the cavities are magnetically dominated (with a plasma-$\beta$ typically well below unity) and contain a significant fraction of the magnetic flux threading the DROD.

\begin{figure*}
    \centering
    \begin{minipage}{0.32\textwidth}
        \includegraphics[width=\textwidth]{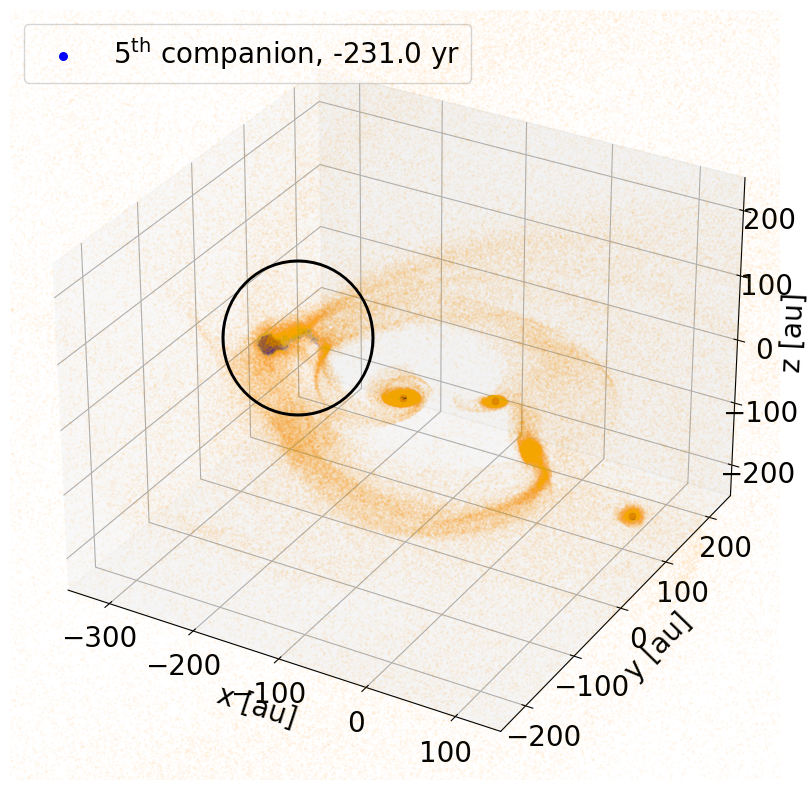}
    \end{minipage}
    \begin{minipage}{0.32\textwidth}
        \includegraphics[width=\textwidth]{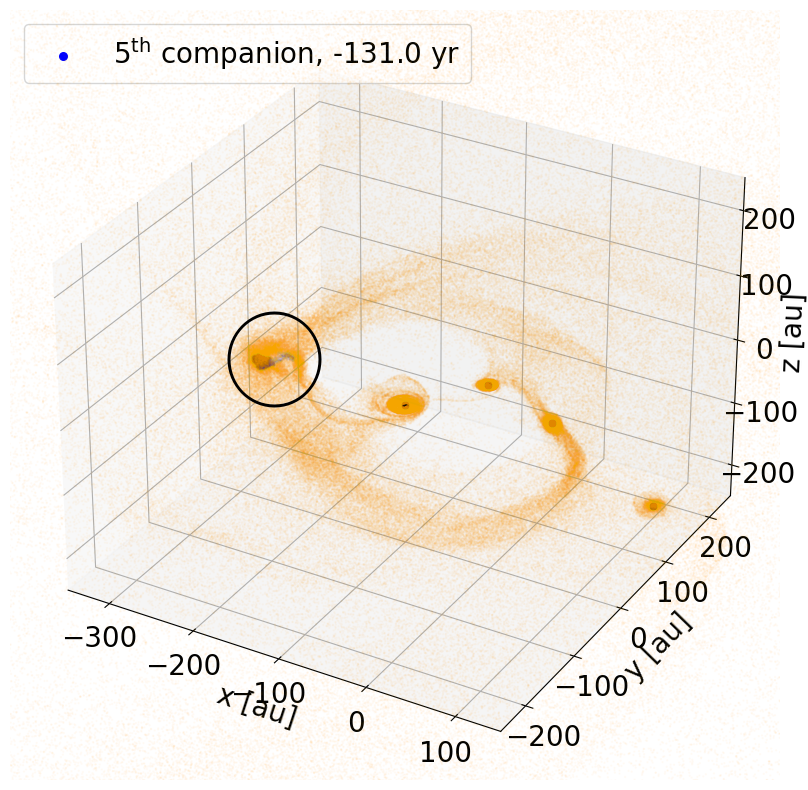}
    \end{minipage}
    \begin{minipage}{0.32\textwidth}
        \includegraphics[width=\textwidth]{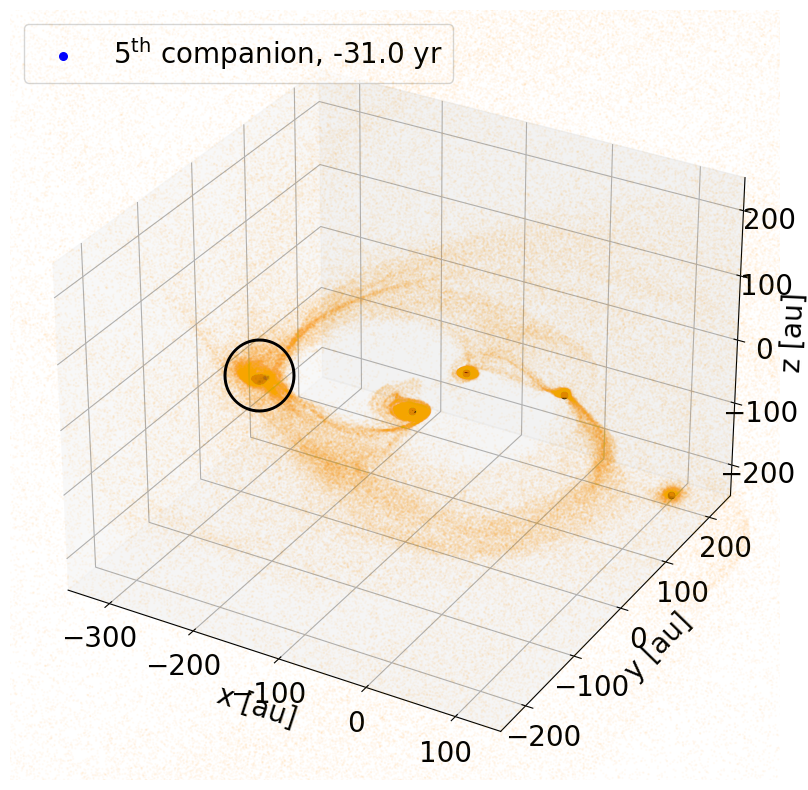}
    \end{minipage}
    \caption{
    3D view of the collisions of dense spiraling filaments leading to the formation of the fifth stellar companion in the reference model. Plotted are Lagrangian tracers, which highlight the colliding filaments leading to the fifth companion's formation. The black circle in each panel highlights the filament collision site. An animation of the figure can be found at \url{https://figshare.com/s/5f173833ca9e9d77cb67}. 
    }
    \label{fig:Collision_3D}
\end{figure*}

To illustrate the collisions leading to stellar companion formation more vividly, we produced a 3D animation highlighting the dense structures in the DROD using the post-processing Lagrangian tracers described in \S~\ref{sec:PostProcessing}. A link to the animation can be found in the caption of Fig.~\ref{fig:Collision_3D}, which shows the evolution of tracers including the three timeframes in the right column of Fig.~\ref{fig:Collision} for the formation of the fifth companion. Figs.~\ref{fig:Collision} and \ref{fig:Collision_3D} and their associated animations leave little doubt that collisions of dense spiraling filaments play a key role in stellar companion formation. Our finding agrees with \citet{Mignon-Risse2021}, who also stresses the importance of filament collisions. 

Similar to the magnetic Toomre parameter $Q_m < 1$ requirement, collision is also a necessary but not sufficient condition for fragmentation. For example, in the less magnetically diffusive model M1.0AD1.0, collisions between dense filaments also happen, yet none of them triggers fragmentation. Because the colliding filaments typically have low values of Toomre-$Q_m$ at the collision sites, these two requirements are often satisfied simultaneously. The lack of fragmentation in Model M1.0AD1.0 indicates it must be missing some ingredient(s) that enabled fragmentation in the more magnetically diffusive reference model.   

\subsection{The Role of Demagnetization in Fragmentation}
\label{sec:condition-pbeta}

\begin{figure*}
    \centering
        \includegraphics[width=\textwidth]{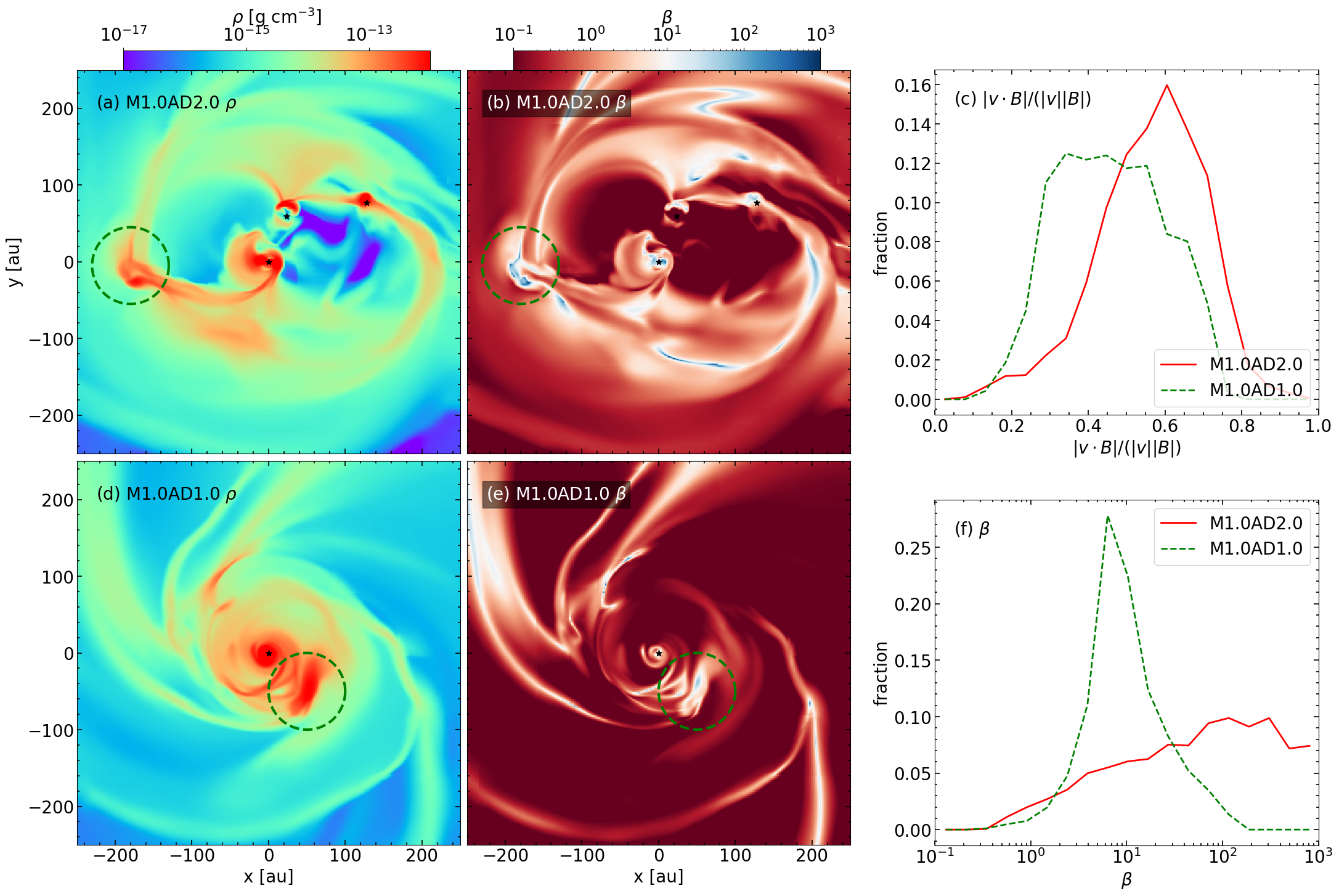}
    \caption{Role of plasma-$\beta$ in fragmentation. Plotted are the distributions of the density and plasma-$\beta$ on the ``midplane'' for Model M1.0AD2.0 (panels [a] and [b], respectively) and M1.0AD1.0 (panels [d] and [e], respectively). The green dashed circle in each panel marks a DROD collision site with significant cross-B field motions, as shown by the distribution of $\cos\theta_\mathrm{ali}$ plotted in panel (c), which is significantly below unity. Panels (b) and (e) show that the collision site in Model M1.0AD2.0 (where fragmentation occurs) is more demagnetized (with a higher plasma-$\beta$) than that in Model M1.0AD1.0, which is further quantified in panel (f), where the distribution of the plasma-$\beta$ is plotted.
    }
    \label{fig:beta}
\end{figure*}

By analyzing the collision sites of the DROD that formed stellar companions in Model M1.0AD2.0 and those that didn't in Model M1.0AD1.0, we find that fragmentation tends to happen when the colliding filaments in the DROD are sufficiently demagnetized, with a plasma-$\beta$ greater than 10 in general. This is illustrated in Fig.~\ref{fig:beta}, which plots the distributions of the density and plasma-$\beta$ on the ``midplane'' (defined in sec.~\ref{sec:ConditionsForFragmentation}) for Models M1.0AD2.0 and M1.0AD1.0 (left and middle panels). The green dashed circle in each panel marks a filament collision site with significant cross-B field motions, as shown by the distribution of $\cos\theta_\mathrm{ali}$ in panel (c). A comparison of panels (b) and (e) shows that the collision site in Model M1.0AD2.0 (where fragmentation occurs) is more demagnetized (with a higher plasma-$\beta$) than that in Model M1.0AD1.0 (where fragmentation does not occur). This difference is quantified in panel (f), where the distribution of the plasma-$\beta$ of the collision site is plotted. Clearly, the collision site of the non-fragmenting M1.0AD1.0 model remains magnetically dominated, with the plasma-$\beta$ mostly below 10. In contrast, the collision site of the fragmenting M1.0AD2.0 model is thermally dominated, with the plasma-$\beta$ mostly above 10. 

%
%
In summary, we find the stellar companions tend to form in highly dynamic dense rotation-dominated structures (DRODs) that have a relatively low magnetic Toomre parameter ($Q_m < 1$), substantial cross-B field motions generated by collisions (corresponding to a relatively low value of the cosine of the alignment angle $\theta_\mathrm{ali}$), and are locally thermally dominated (with a plasma-$\beta \gtrsim 10$). These tendencies stress the importance of the non-ideal MHD effects in the formation of multiple stellar systems in molecular cloud cores with dynamically significant magnetic fields. 

\section{Discussion}
\label{sec:discussion}

\subsection{Close Misaligned Companion Formation Through DROD Fragmentation}
\label{sec:nonDiskFrag}

As mentioned in \S~\ref{sec:introduction}, disk fragmentation is a widely discussed channel for forming close binary/multiple systems \citep[e.g.][]{Offner2023}. It is thought to happen, in particular, during the deeply embedded phase of star formation, when the mass accretion from the protostellar envelope onto the disk exceeds the mass accretion rate through the disk onto the central protostar \citep[e.g.][]{Kratter2010}, with the resultant mass accumulation driving the disk fragmentation and stellar companion formation. Two generic expectations for this scenario are: (1) spiral arms develop in the gravitationally unstable disk and dominate the disk angular momentum transport, and (2) the circumstellar disk of the companion, if any, should be aligned with that of the primary. In this subsection, we demonstrate that these expectations are not met in our simulations, pointing to a close binary/multiple formation mechanism related to, but distinct from, the traditional disk fragmentation. 

\subsubsection{Magnetically Dominated Angular Momentum Transport in DRODs}
\label{sec:magTransportAngMomentum}

\begin{figure}
    \centering
    \includegraphics[width=\columnwidth]{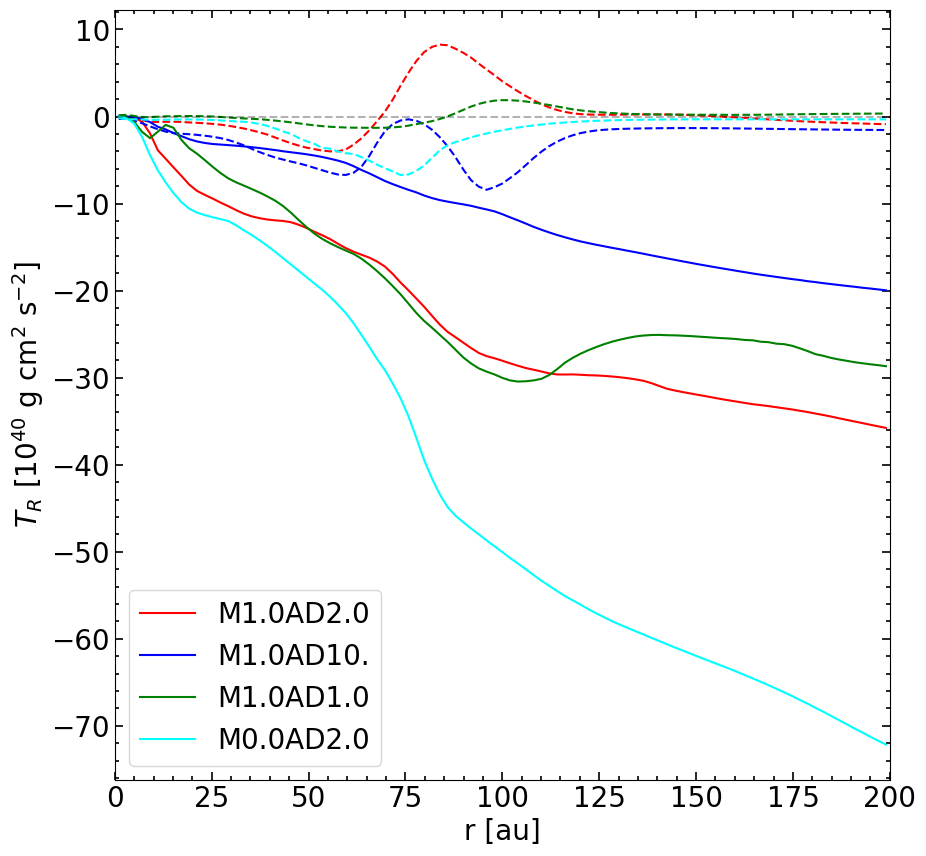}
    \caption{Cumulative angular momentum transport rates by gravitational (dashed lines) and magnetic torques (solid) for different models. The magnetic torque dominates angular momentum transport in the M1.0AD2.0, M1.0AD1.0, and M1.0AD2.0 models. In the M1.0AD10 model, gravity and magnetic fields transport angular momentum at comparable rates within the innermost 100 au.}
    \label{fig:torque}
\end{figure}

In our simulations, prominent spiraling filaments are present in both the large-scale gravo-magneto-sheetlets and the small-scale DRODs (see, e.g., Fig.~\ref{fig:sheetlet_3D} and the lower-left panel of Fig.~\ref{fig:Q}). The larger-scale spirals in the collapsing sheetlets are produced by the inhomogeneities induced by the initial turbulence unrelated to gravitational instability. The smaller-scale dense spiraling filaments in the DRODs are often spatially connected to the larger-scale spirals in the sheetlets and are thus likely resulting from non-uniform feeding by the sheetlets rather than the need to transport angular momentum from large to small distances. Indeed, the gravitational torque due to the spiraling filaments is substantially smaller than the magnetic torque on the scale of the DROD in the reference model, as illustrated by the red lines in Fig.~\ref{fig:torque}, where the accumulative torques up to a given radius from the primary are plotted as a function of radius.
The accumulative torque is computed by summing the total gravitational or magnetic torque on the gas within 50 au above or below the midplane (defined in sec.~\ref{sec:Q}), i.e.
\begin{equation}
    T_R(R)=\int_0^{2\pi}\int_{-z}^z\int_0^R f_\phi \ R dV,
\end{equation}
where $f_\phi$ is the force in the $\phi$ direction in the cylindrical coordinate with $z$-axis perpendicular to the midplane and center on the primary; $R$ is the cylindrical radius from the primary.
For the models with more than one star formed, the time plotted is chosen to be within 100 years before the secondary formation. Given the overwhelming domination of the gravitational torque by the magnetic torque in the reference model (M1.0AD2.0), it is reasonable to surmise that the spiraling filaments develop in the DROD of this model {\it not} to transport angular momentum as in the traditional disk fragmentation picture, but as an inheritance from the larger-scale sheetlets that already contain prominent spirals. This difference may have contributed, at least in part, to the frequent collisions of the dense spiraling filaments in the reference model, which are key to stellar companion formation. 

In the non-turbulent model M0.0AD2.0 and less magnetically diffusive model M1.0AD1.0, the magnetic torque also dominates over the gravitational torque, similar to the reference model. However, in the more magnetically diffusive model M1.0AD10, the gravitational torque can be comparable to the magnetic torque within the innermost 100 au. Because of the high diffusivity in the M1.0AD10 model, the magnetic field is expected to be weaker than in the other models, so gravity is expected to be more important. It is, therefore, unsurprising that the secondary formation in the M1.0AD10 model is closer to the traditional view of disk fragmentation in the absence of magnetic fields. Without less magnetic pressure hindering the collapse of self-gravitating gas (see sec.~\ref{sec:ConditionsForFragmentation}), the secondary formed earlier in this case than in other models (fig.~\ref{fig:mass_summary}).

\subsubsection{Circumstellar Disk Misalignment}
\label{sec:misalignment}

\begin{figure*}
    \centering
    \includegraphics[width=\textwidth]{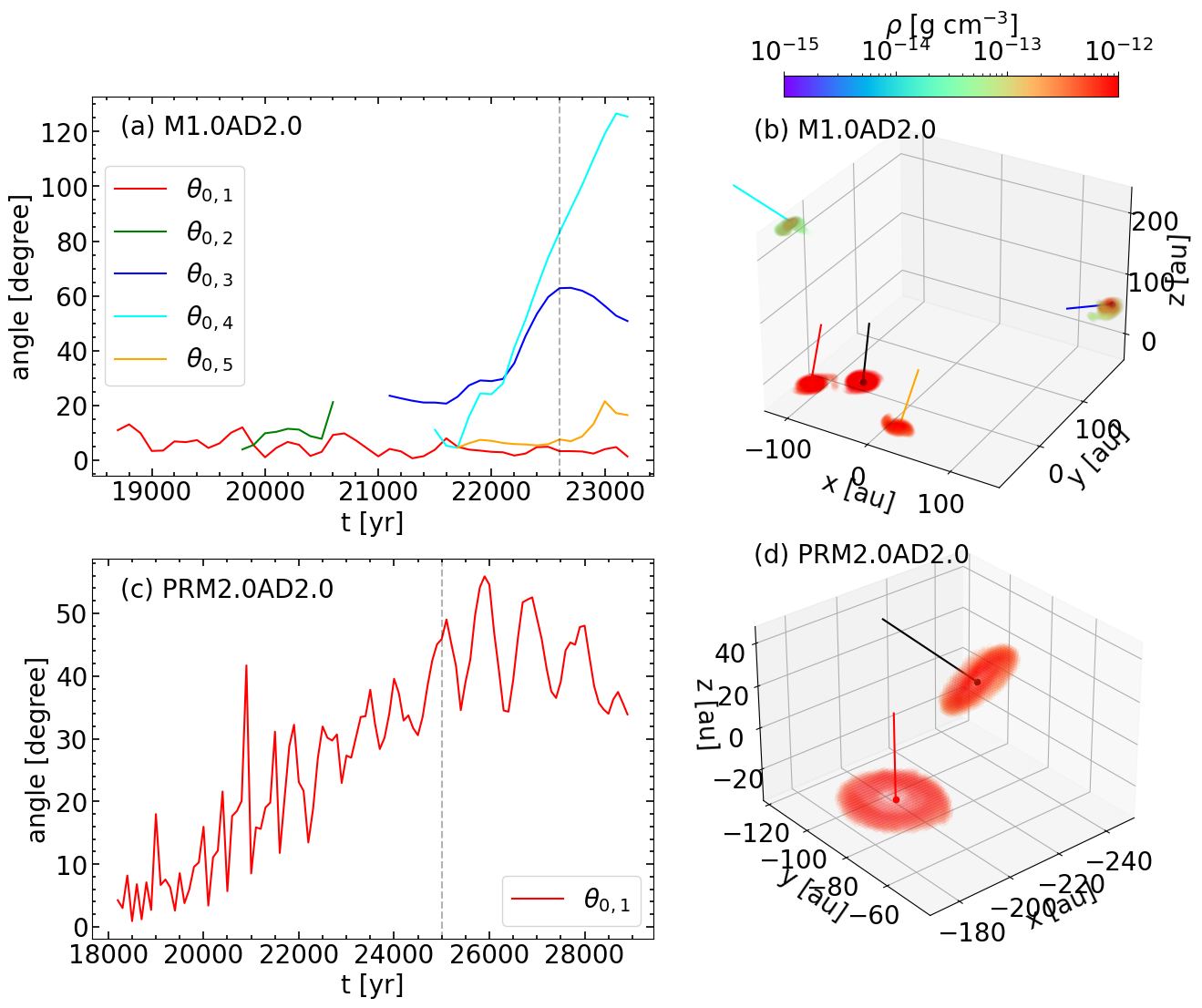}
    \caption{Circumstellar disk misalignment in Models M1.0AD2.0 (upper panels) and PRM2.0AD2.0 (lower panels). The left panels ([a] and [c]) plot the evolution of the angles between the circumstellar disk axes of stellar companions and that of the primary, showing disk misalignment, particularly at late times. The right panels ([b] and [d]) show a visualization of the circumstellar disk misalignment at the time marked by the vertical dashed line in the corresponding left panel.  An animated version of the figure can be found at \url{https://figshare.com/s/f4fb7069325dad9d07de}. 
    }
    \label{fig:disk_misalignment}
\end{figure*}


Even more intriguing is the misalignment between the stellar systems, which is not expected in the simplest version of the traditional disk fragmentation scenario. We will demonstrate the misalignment using two simulations: the reference run, where several stellar objects are formed with moderately misaligned disks, and a new simulation (Model PRM2.0AD2.0 in Table~1), where a binary system is formed with a more pronounced disk misalignment. 

We will use the angular momentum axis of the circumstellar material within $25$~AU of each stellar object as a proxy for the circumstellar disk axis because the disks are relatively small and may not be adequately resolved. The results are shown in Fig.~\ref{fig:disk_misalignment}, which plots the evolution of the angles between the circumstellar disk axes of the stellar companions and that of the primary for the reference model (M1.0AD2.0, panel [a]) and Model PRM2.0AD2.0 (panel [c]). Both cases show considerable disk misalignment, particularly at relatively late times, as illustrated in the right panels ([b] and [d]), which show a visualization of the circumstellar disk misalignment at the time marked by the vertical dashed line in the corresponding left panel. 

Note that the misalignment angle takes a value between $0$ and $180$ degrees because disks can rotate in opposite directions. In the reference model, the disk axis of the first companion (particle 1) remains aligned rather closely with that of the primary. It is not true for other companions, especially the third and fourth companions. The disk axis of the third companion starts with a 25-degree misalignment from that of the primary disk and increases to as large as $60$ degrees. The disk misalignment angle of the fourth companion starts around 15 degrees but quickly increases above 90 degrees, reaching $\sim 120$~degrees towards the end of the simulation. The counter-rotation is an extreme example of the deviation from the traditional disk fragmentation scenario for companion formation. The reason for the disk misalignment is that the stellar companions in our core-collapse simulations with magnetic fields and turbulence are formed in highly structured DRODs that are much more dynamically active than the circum-primary disk envisioned in the traditional disk fragmentation scenario. The DRODs are fed from different directions by the larger-scale warped gravo-magneto-sheetlets with directions of angular momentum different from that of the primary disk. It is more so at a later time when the feeding sheetlet material originates from a larger initial distance, where it is expected to be more strongly perturbed by turbulence. 
 
To illustrate the above picture more clearly, we consider a more perturbed version of the reference run with a stronger initial turbulence of rms Mach number of 2 and an initial magnetic field (along $x$-direction) perpendicular to the rotation axis (along $z$-direction; Model PRM2.0AD2.0 in Table~\ref{tab:para}). Only a single companion is formed in this simulation, with a disk axis that starts close to that of the primary but becomes increasingly more misaligned later, reaching $\sim 40^\circ$ at the end of the simulation (see the lower panels of Fig.~\ref{fig:disk_misalignment}). The periodic variation in the inclination angle reflects the orbiting of the secondary around the primary. This example reinforced the notion that the collapse of turbulent magnetized cloud cores can naturally lead to close misaligned stellar systems. Our findings agree with the SupAS model of \cite{Mignon-Risse2021}, which also produced misaligned close multiple systems (not shown in the paper, private communication).

\subsubsection{A Variant of the Disk Fragmentation Mechanism}
\label{sec:variant}

\begin{figure}
    \centering
    \includegraphics[width=\columnwidth]{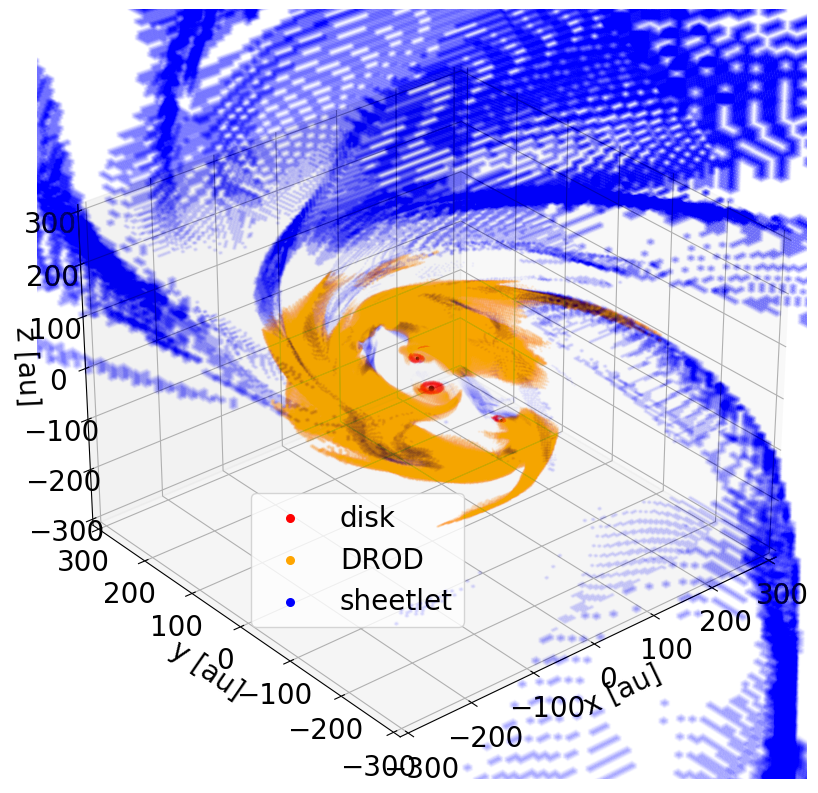}
    \caption{Key components of the DROD fragmentation mechanism for binary/multiple formation. The circumstellar disk around each star is shown in red. The irregular, orange-colored structure is the DROD, which is fed by larger-scale sheetlets (shown in blue). An animated version of this figure can be found at \url{https://figshare.com/s/9b8238b28694a8aab8ca}.}
    \label{fig:disk_DROD_sheetlet}
\end{figure}

The circumstellar disk misalignment and magnetic domination of the angular momentum transport motivate us to propose a variant to the traditional disk fragmentation -- DROD fragmentation. The key components of this new variant are illustrated in Fig.~\ref{fig:disk_DROD_sheetlet}, which shows that the DROD (orange-colored, defined in sec.~\ref{sec:RotationPileUp}) is fed magnetized material by larger-scale warped gravo-magneto-sheetlets (blue-colored, defined in sec.~\ref{sec:sheetlets}) from different directions. The resulting DROD contains dense spiraling filaments that do not generate enough gravitational torque to dominate the angular momentum transport. The filaments collide with one another to create conditions conducive to fragmentation and stellar companion formation. This more dynamic variant allows for the {\it in-situ} formation of misaligned systems on the 100-au scale.  

\subsection{Connection to Observations}
\label{sec:Obs}


The DRODs and associated features discussed in the preceding sections can be used to interpret observations of binary/multiple protostellar systems. For illustration purposes, we will focus on three representative systems: L1448 IRS3B, NGC 1333 IRAS2A, and VLA1623.  


\begin{figure}
    \includegraphics[width=\columnwidth]{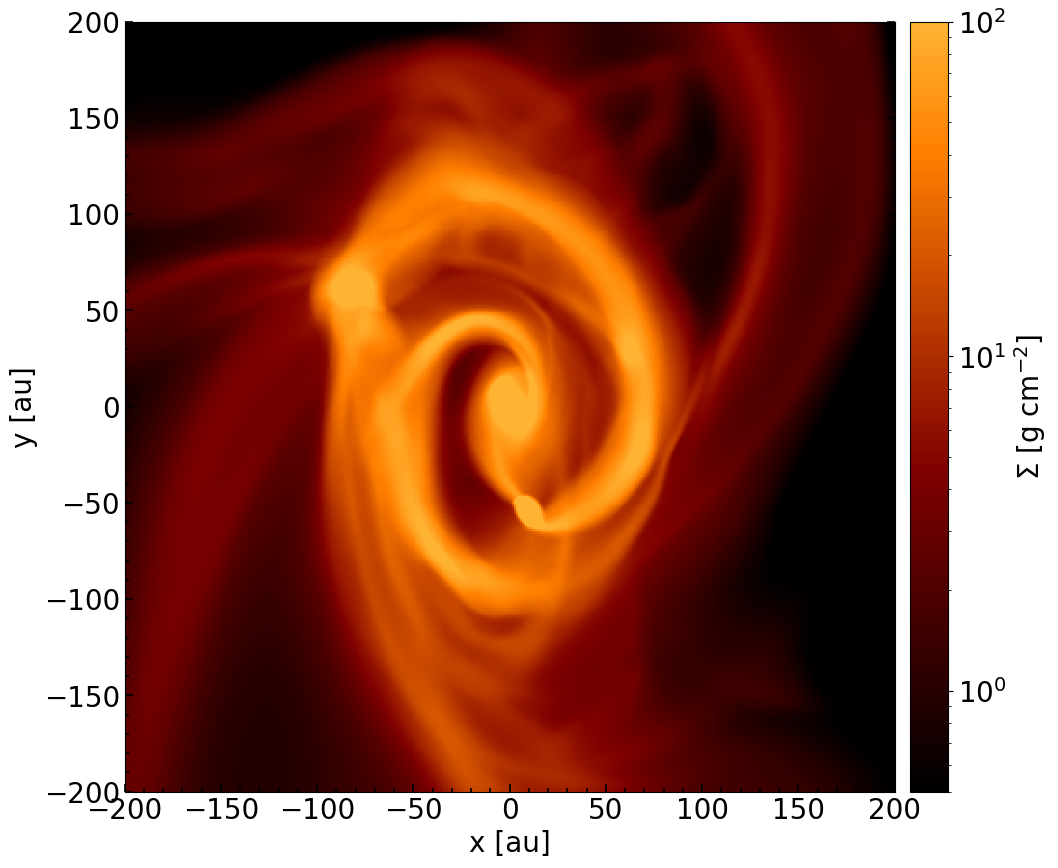}
    \caption{A simulated triple system analogous to L1448 IRS3B, but formed through collisions of dense rotationally dominated structures (DRODs). The figure is the column density in the M1.0AD2.0 model.}
    \label{fig:IRS3B}
\end{figure}

L1448 IRS3B is a Class 0 triple system considered to be the prototype of multiple systems formed through disk fragmentation \citep{Tobin2016b}.  Two stellar objects have already formed, and a third object appears to be caught in formation in the circumbinary disk of the binary. A key piece of evidence for disk fragmentation was that the Toomre Q parameter was estimated to be close to unity near the tertiary \citep{Reynolds2021}. However, a low Toomre-Q value is not definitive proof that the stellar companion formed through the traditional disk fragmentation, where spiral arms necessitated by the need to redistribute disk angular momentum become self-gravitating and collapse gravitationally. Our reference simulation (Model M1.0AD2.0) passes through a phase that resembles the observed system, with two stars already formed and a third on its way, as shown in Fig.~\ref{fig:IRS3B}. In our case, the third stellar object (the second stellar companion) was formed through collisions between different streams of dense material with Toomre Q below unity (see, e.g., Fig.~\ref{fig:Q}e) that are fed by, and spatially connected to, the larger scale collapsing gravo-magneto-sheetlets. 

NGC 1333 IRAS2A is a sub-arcsecond Class 0 protobinary system (separated by $\sim 143$~au) studied in detail by \citet{Tobin2015, Tobin2016} as part of the VLA/ALMA Nascent Disk and Multiplicity (VANDAM) survey. A salient feature of this system is that the two protostars drive two nearly orthogonal CO outflows (see their Fig.~6). The large misalignment led to the suggestion that the two stellar components were formed independently at a much larger initial distance through turbulent fragmentation and subsequently moved closer together, as found in some turbulent fragmentation simulations \citep[e.g.,][]{Offner2010}. However, we have demonstrated that highly misaligned systems on the 100-au scale can be formed {\it in situ} (i.e., without the need for orbital tightening by a large factor) from dense rotationally dominated structures that are much more disorganized and dynamically active than the traditional disks (see, e.g., Fig.~\ref{fig:disk_misalignment}d). The larger-scale warped sheetlets deliver materials with different angular momentum directions to the stellar companions, which is key to producing and maintaining the misalignment. 

An even more extreme case of misalignment is VLA1623, the first Class 0 object observed and a prototype of the class \citep{Andre1993}. It is a deeply embedded quadruple system consisting of a close binary pair (VLA1623A1 and VLA1623A2), a relatively close companion with $\sim100$~au separation (VLA1623B) and another companion much farther away (VLA1623W). The disk of component "B" is highly inclined with respect to the circumbinary disk around components "A1" and "A2," and rotates in a direction opposite to that of the circumbinary disk \citep[see][their Fig.~21]{Ohashi2022}. It is unlikely to have formed through traditional disk fragmentation. A clue for its origin may come from the spatially and kinematically distinct accretion streams identified by \citet[][see their Fig.~19; {see also \citealp{Codella2024}}]{Hsieh2020}, which may have formed and maintained the highly misaligned system. It is plausible that the accretion streams correspond to the collapsing sheetlets in our picture, although magnetic field information and more detailed analysis are needed to firm up the identification\footnote{We note that dense filaments are produced in many star and disk formation simulations including magnetic fields and turbulence \citep[see, e.g.,][]{Kuffmeier2019}. As discussed by \citet[][see their introduction and sec.~7.1, and references therein]{Tu2023}, many, if not all, of them are likely produced by the interplay between magnetic fields, turbulence, and gravity, as is the case for our sheetlets, although detailed analysis of their magnetic fields and dynamics is needed to be sure.} In any case, this protostellar multiple system presents a challenge to the traditional disk fragmentation scenario but could be produced through the more dynamic scenario of DROD fragmentation proposed here, which can produce counter-rotating circumstellar structures (see the cyan line in Fig.~\ref{fig:disk_misalignment}a, which shows that the misalignment angle for the fourth stellar companion goes beyond $90^\circ$ at late times). 

\section{Conclusions}
\label{sec:conclusion}

We have carried out a set of non-ideal MHD simulations of binary/multiple formation from the gravitational collapse of magnetized, turbulent, low-mass molecular cloud cores. Our main conclusions are as follows. 

\begin{enumerate}
    \item As in the case of single star formation in turbulent, magnetized cores, the inner protostellar envelope is dominated by dense gravo-magneto-sheetlets, which are a turbulence-warped version of the classic pseuodisk resulting from the anisotropic magnetic resistance to the gravitational collapse. These sheetlets feed mass, magnetic fields, and angular momentum to a dense rotation-dominated structure -- a DROD -- where fragmentation leads to binary/multiple star formation. We define a DROD to be the structure where density is larger than $3\times10^{-15}\ \mathrm{g\ cm^{-3}}$ and rotating at $>0.9\ \times$ local keplerian speed (computed using the enclosed mass).

    \item The DROD is a more dynamic version of the traditional disk with dense spiraling filaments created by inhomogeneous feeding from the highly structured larger-scale sheetlets rather than the need for angular momentum transport, which is dominated by magnetic braking. Collisions between the dense rotation-dominated filaments play a key role in pushing the local magnetic Toomre parameter $Q_\mathrm{m}$ below unity at the collision sites, leading to gravitational collapse and stellar companion formation provided that the local material is sufficiently demagnetized, with a plasma-$\beta$ of 10 or more. This DROD fragmentation scenario for binary/multiple star formation is a more dynamic variant of the traditional disk fragmentation in a turbulent, magnetized cloud core. It can naturally produce {\it in situ} misaligned systems on the 100-au scale, often detected with high-resolution ALMA observations. 

    \item Our simulations highlight the importance of non-ideal MHD effects in binary/multiple star formation. For our modeled cloud cores, the fragmentation is completely suppressed when the magnetic diffusivity is relatively low (e.g., Model M1.0AD1.0). Fragmentation occurs when the magnetic diffusivity is large enough, with the stellar masses and orbital parameters strongly affected by the diffusivity level (compare, e.g., Model M1.0AD2.0 and M1.0AD10). The magnetic diffusivity depends on the cosmic ray ionization rate and grain size distribution, which are uncertain. The uncertainties make it challenging to develop a complete theory for binary/multiple star formation in magnetized clouds. 
\end{enumerate}

\section*{Acknowledgements}

We thank the referee, Dr.~Rapha\"el Mignon-Risse, for a detailed and constructive report. This work is supported by NASA 80NSSC18K1095, NASA 80NSSC20K0533, NSF AST-1910106, and NSF AST-2307199. Computing resources were provided by the NASA High-End Computing (HEC) Program through the NASA Advanced Supercomputing (NAS) Division at Ames Research Center and the RIVANNA supercomputer at the University of Virginia. YT acknowledges support from an interdisciplinary fellowship from the University of Virginia. 

\section*{Data Availability}

The data underlying this article will be shared on reasonable request to the corresponding author.



\bibliographystyle{mnras}
\bibliography{example} 





\appendix
\section{3D Visualization with Unity}

To better visualize the formation of a multiple-star system, we use the Unity Engine \citep{Unity}, a 3D game engine, to generate movie-like videos of our simulation. 

We visualize the reference model (M1.0AD2.0 model). The gas is visualized with spheres, with each sphere representing the same gas mass. The colors of the spheres follow the rainbow color scheme (similar to panel a of fig.~\ref{fig:beta}), representing the gas density at their locations. Blue means lower density and red means higher density. The gas particle trajectories are calculated with the post-process Lagrangian tracer particle code described in sec.~\ref{sec:PostProcessing}. The sizes of the stars are proportional to their masses. To account for the stellar illumination on gas for artistic purposes, each star is accompanied by a white light source at its location after its formation.
For artistic purposes, the post-process Lagrangian simulation is modified in two ways. First, to better visualize the disk and minimize the impact of the relatively low MHD output frame rate, we force the gas particle to orbit in Keplerian orbit around a star if the particle is in the vicinity of the star (typically within $25$au from the star) and the particle orbital velocity is $>0.7$ local Keplerian velocity relative to its host star. Second, 
the second companion is not shown in the movie because it merged with the first companion. 

\begin{figure*}
    \centering
    \includegraphics[width=\textwidth]{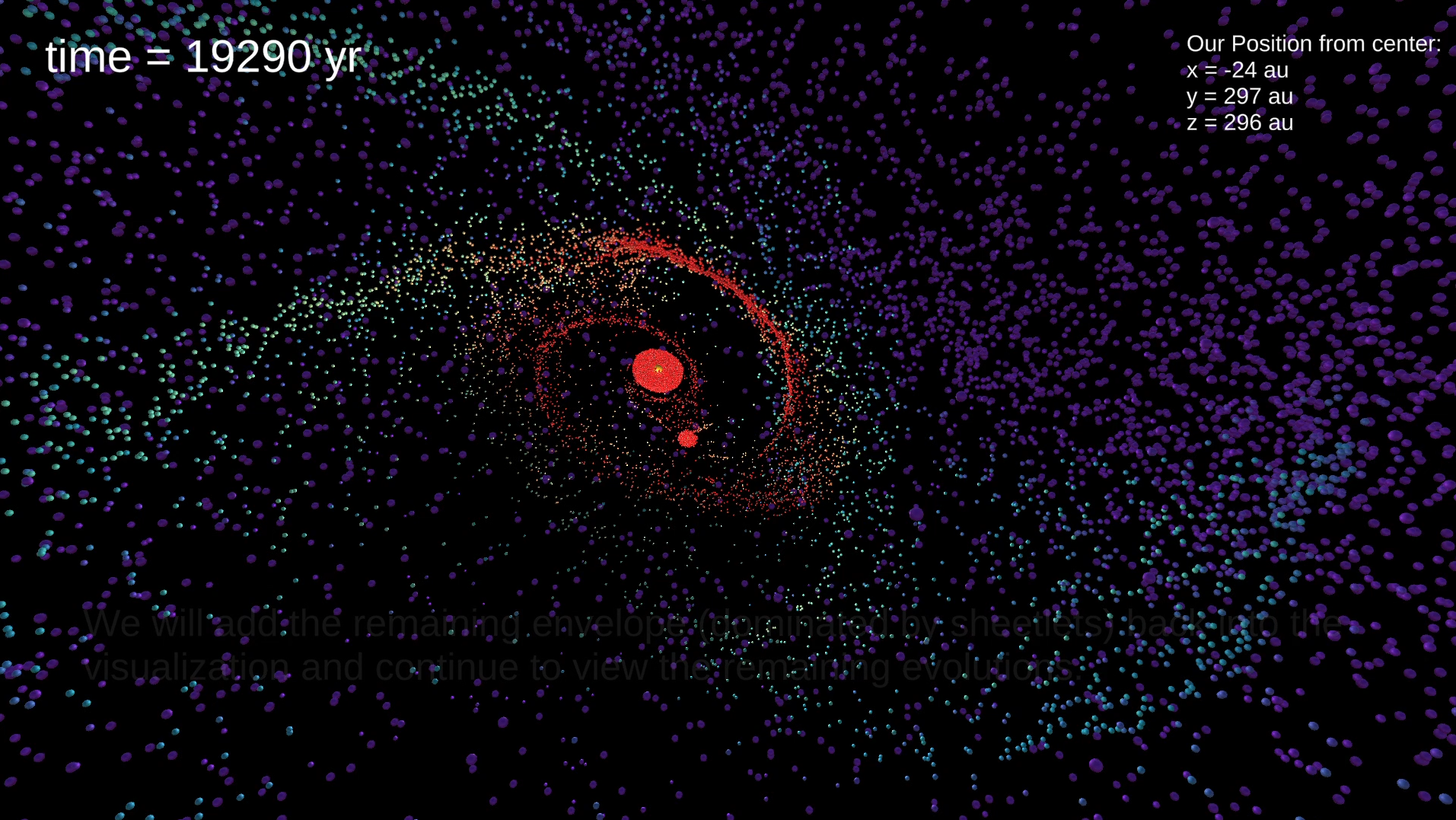}
    \caption{One representative frame of the artististic rendering of the M1.0AD2.0 model with Unity \citep{Unity}. The movie can be found on YouTube at \url{https://www.youtube.com/watch?v=s-eHv7cAObw}}
    \label{fig:Unity}
\end{figure*}

The movie can be on YouTube at \url{https://www.youtube.com/watch?v=s-eHv7cAObw}. We show a representative frame of the movie in fig.~\ref{fig:Unity}.

\bsp	
\label{lastpage}
\end{document}